\documentclass[a4paper]{iopart}
\pdfoutput=1
\usepackage[utf8]{inputenc}
\usepackage{amsfonts,amsmath,amssymb,graphicx,epstopdf,verbatim}
\usepackage{rotating}
\usepackage{epstopdf}

\usepackage{xcolor}
\renewcommand{\textcolor}[2]{#2}

\usepackage[colorlinks]{hyperref}
\definecolor{darkred}{rgb}{0.5,0,0}
\definecolor{darkgreen}{rgb}{0,0.5,0}
\definecolor{darkblue}{rgb}{0,0,0.5}
\definecolor{pastelBlue}{HTML}{AECBFA}     
\definecolor{tensorGreen}{HTML}{81C784}  
\definecolor{tensorOrange}{HTML}{FFB74D} 
\definecolor{tensorTeal}{HTML}{43B8AC} 
\definecolor{pastelMarron}{RGB}{210, 180, 140} 
\definecolor{pastelPink}{HTML}{F8BBD0}     
\definecolor{pastelYellow}{HTML}{FFF59D}   
\definecolor{pastelCyan}{HTML}{B2EBF2}     
\definecolor{pastelRed}{HTML}{FFABAB} 
\definecolor{pastelCoral}{HTML}{FFCCBC}    
\hypersetup{colorlinks,
linkcolor=darkred,
filecolor=darkgreen,
urlcolor=darkblue,
citecolor=darkgreen}
\usepackage{braket}
\usepackage{overpic}
\usepackage[square,numbers]{natbib}
\usepackage{orcidlink}
\usepackage{tikz}
\usetikzlibrary{positioning,shapes.geometric,decorations.pathreplacing,calc}

\newcommand{\nep}{\operatorname{e}}

\tikzset{
  tensorCir/.style={
    circle,
    draw=black,
    line width=2pt, 
    minimum size=0.75cm, 
    font=\large
  },
  tensorSqua/.style={
    rounded corners=4pt,
    draw=black,
    line width=2pt,
    minimum size=0.6cm
  },
  tensorTri/.style={
    regular polygon,
    regular polygon sides=3,
    draw=black,
    line width=2pt,
    minimum size=1cm,
    transform shape,
    font=\small,
    inner sep=0.5pt
  },
  leg/.style={
    line width=1.75pt
  },
  index/.style={
    line width=1.75pt 
  }
}

\begin{document}

\title[Operator delocalization]{Operator delocalization in disordered spin chains via exact MPO marginals}

\author{Jonnathan Pineda-Jiménez~\orcidlink{0009-0008-8180-3089}}
\address{International School for Advanced Studies (SISSA), Via Bonomea, 265, 34136 Trieste, Italy}
\vspace{-0.1cm}
\author{Mario Collura~\orcidlink{0000-0003-2615-8140}}
\address{International School for Advanced Studies (SISSA), Via Bonomea, 265, 34136 Trieste, Italy}
\address{INFN, Sezione di Trieste, Via Valerio 2, 34127 Trieste, Italy}
\vspace{-0.1cm}
\author{Gianluca Passarelli~\orcidlink{0000-0002-3292-0034}}
\address{Dipartimento di Fisica ``E. Pancini'', Università di Napoli Federico II, I-80126 Napoli, Italy}
\vspace{-0.1cm}
\author{Procolo Lucignano~\orcidlink{0000-0003-2784-8485}}
\address{Dipartimento di Fisica ``E. Pancini'', Università di Napoli Federico II, I-80126 Napoli, Italy}
\vspace{-0.1cm}
\author{Davide Rossini~\orcidlink{0000-0002-9222-1913}}
\address{Dipartimento di Fisica dell’Università di Pisa and INFN, Largo Pontecorvo 3, I-56127 Pisa, Italy}
\vspace{-0.1cm}
\author{Angelo Russomanno~\orcidlink{0009-0000-1923-370X}}
\address{Dipartimento di Fisica ``E. Pancini'', Università di Napoli Federico II, I-80126 Napoli, Italy}
\vspace{-0.1cm}
\begin{abstract}
We investigate operator delocalization in disordered one-dimensional spin chains by introducing -- besides the already known operator mass -- a complementary measure of operator complexity: the operator length. Like the operator nonstabilizerness, both these quantities are defined from the expansion of time-evolved operators in the Pauli basis. They characterize, respectively, the number of sites on which an operator acts nontrivially and the spatial extent of its support. We show that both the operator mass and length can be computed efficiently and exactly within a matrix-product-\textcolor{red}{state} (MPS) framework, providing direct access to their full probability distributions, without resorting to stochastic sampling. Applying this approach to the disordered XXZ spin-1/2 chain, we find sharply distinct behaviors in non-interacting and interacting regimes. In the Anderson-localized case, operator mass, length, and operator entanglement entropy rapidly saturate, signaling the absence of scrambling. By contrast, in the many-body localized (MBL) regime, for arbitrarily weak interactions, all quantities exhibit a robust logarithmic growth in time, consistent with the known logarithmic light cone of quantum-correlation propagation in MBL. We demonstrate that this behavior is quantitatively captured by an effective $\ell$-bit model and persists across system sizes accessible via tensor-network simulations. 
\end{abstract}

\section{Introduction}
%
  Quantum thermalization is distinct from classical thermalization, as it entails the development of entanglement and its progressive spreading over increasingly delocalized degrees of freedom~\cite{SciPostPhysLectNotes.20}. This statement is made more quantitative by looking at the operator properties. In a thermalizing dynamics, operators become progressively more delocalized: as one can see by applying the Baker-Campbell-Hausdorff formula, they evolve into increasingly extended operator strings, with higher-order nested commutators becoming more relevant at later times. Consequently, information is dispersed into increasingly nonlocal degrees of freedom, a process known as scrambling~\cite{Swingle}.
  
  To quantify scrambling, the out-of-time-order correlator (OTOC) has been introduced~\cite{Larkin_69, black}. It is defined as $F(t)=\braket{\hat{W}^\dagger(t) \, \hat{V}^\dagger \, \hat{W}(t) \, \hat{V}}$, where $\hat{V}$ and $\hat{W}$ are local operators whose supports lie at a given distance, and $\hat{W}(t)=\nep^{i\hat{H} t}\hat{W}\nep^{-i\hat{H} t}$ is the time-evolved version with some local Hamiltonian $\hat{H}$ (hereafter we work in units of $\hbar=1$). The OTOC displays a propagating wavefront: ahead of the front it grows rapidly, typically exponentially, whereas behind it saturates to a constant value~\cite{swing_nat,Swingle}. A ballistic propagation of the wave front has been observed in strongly chaotic holographic and Sachdev-Ye-Kitaev coupled systems~\cite{black,Gu_2017,Song_2017,Ben_Zion_2018} and in the $O(N)$ model~\cite{PhysRevD.96.065005} and weakly interacting diffusive metals~\cite{PhysRevX.7.031047} near the wavefront. On the other hand, a diffusive propagation emerges in chaotic quantum circuits~\cite{PhysRevX.7.031016,PhysRevX.8.021014,PhysRevX.8.031057,PhysRevX.8.031058}, while a subdiffusive propagation occurs in integrable delocalized quantum systems~\cite{swing_nat}. In the last case, the system does not thermalize, as it reaches a generalized-Gibbs-ensemble steady state that is nonthermal~\cite{Alba_2017,Alba_2018}, although quantum information is anyway scrambled in nonlocal degrees of freedom.
  
  In disordered interacting systems, the OTOC displays a distinctive behavior, since its wavefront propagates as a logarithmic light cone in time~\cite{PhysRevB.95.060201,FAN2017707,chen2016universallogarithmicscramblingbody,Huang_2016}. Such logarithmically spreading fronts are a common feature of strongly disordered systems: they appear in the logarithmic growth of the half-chain entanglement entropy~\cite{PhysRevB.77.064426,PhysRevLett.109.017202,PhysRevLett.110.260601,De_Chiara_2006} (the entanglement entropy is a measure of an entanglement correlation length displaying a logarithmic light cone~\cite{Deng_2017}), as well as in heat transport when one side of the system is coupled to a thermal bath~\cite{PhysRevB.110.134204}. More generally, these phenomena are associated with a logarithmic propagation bound~\cite{kim2014localintegralsmotionlogarithmic,logor,toniolo2024stabilityslowhamiltoniandynamics}, that is the disordered counterpart of the ballistic Lieb–Robinson bound which is valid in clean systems.
  This slow, logarithmic propagation is also related
  to hindered thermalization. Indeed, such systems are expected not to thermalize at infinite times, a phenomenon known as many-body localization (MBL)~\cite{RevModPhys.91.021001}. While the ultimate absence of thermalization in the thermodynamic and long-time limits remains under debate~\cite{PhysRevLett.130.250405,PhysRevLett.133.116502,PhysRevE.102.062144,PhysRevB.102.064207,PhysRevE.104.054105,PhysRevB.105.174205, Sierant_2022_Challeng, Sierant_2025_MBL_Class}, these systems nevertheless relax extremely slowly. This agrees with the fact that the scrambling of quantum information is also exceptionally slow, being characterized by a logarithmic light-cone propagation.
  An efficient evaluation of the OTOC in terms of matrix-product operators (MPO) has been proposed~\cite{swing_nat} and proved itself to be highly effective in accessing the OTOC slightly ahead of the propagation front, thereby providing valuable information on the wavefront dynamics. We emphasize that the OTOC provides a space-time information on the evolution of the scrambling, without reference to any particular operator basis.
  
  In recent years, a particular operator basis, namely, the Pauli-string basis, has attracted attention in the literature. Its relevance stems from the fact that delocalization in this basis of an evolving quantum state---evaluated through the stabilizer R\'enyi entropy (SRE)~\cite{PhysRevLett.128.050402,PhysRevA.106.042426}---provides a measure of the distance from the set of stabilizer states, which can be efficiently prepared using classical resources~\cite{PhysRevA.57.127,gott}. In strongly disordered spin chains the dynamics of the SREs has been only investigated at the level of the states in strongly disordered spin chains~\cite{Falcao_2025_SREMBL,li_2026}, while delocalization of operators in the Pauli basis has been recently considered in the context of open quantum systems~\cite{PhysRevLett.131.160402}. 
  Each Pauli string can be assigned a ``mass", defined as the number of non-identity Pauli matrices it contains. The mass of a time-evolving operator is then obtained by averaging this Pauli-string mass over the probability distribution given by the squared scalar products with the Pauli-string basis. The resulting mass quantifies the degree of delocalization of the operator.
  
  Building on this idea, we introduce a length associated with an operator, based on its expansion in the Pauli operators. Specifically, in addition to the mass, each Pauli-string operator is assigned a length, given by the position of the rightmost site differing from the identity. The length of a time-evolving operator is then obtained by averaging this Pauli-string quantity over the distribution defined by the squared scalar products.
  The evolution of such quantities provides a complementary perspective on the scrambling: as the operator delocalizes, it increasingly overlaps with more nonlocal Pauli strings, causing its length and mass to increase over time.
  This leads to a crucial advantage: in fact, both the operator length and the operator mass can be computed efficiently and exactly within the MPO representation of the time-evolved operator. Our method provides direct access to the full probability distribution of these quantities without any need for stochastic sampling, in contrast with approaches based on full operator tomography via perfect sampling of the Pauli spectrum, which are inherently probabilistic and whose required number of samples may scale exponentially with the system size. Although MPO-based simulations generally require truncation to remain computationally feasible, we find that, for an Anderson-localized integrable dynamics, no truncation is needed, while in MBL systems, the required bond dimension increases only linearly in time, reflecting the logarithmic increase in time we observe in the operator entanglement entropy~\cite{bandyopadhyay2005entanglingpowerquantumchaotic,PhysRevA.76.032316,PhysRevB.79.184416} of an initially localized operator, and ensuring that the computation remains polynomially efficient.
  
  We focus on the MBL case and find that the length of an initially localized operator increases logarithmically in time. Looking at the scrambling from this perspective, the results are in full agreement with the OTOC wavefront propagating with a logarithmic light cone, as found in Refs.~\cite{PhysRevB.95.060201,FAN2017707,chen2016universallogarithmicscramblingbody,Huang_2016}. Following this literature, we discuss a simple $\ell$-bit model based on localized integrals of motion that provides the same logarithmic increase in time of the operator length~\cite{PhysRevB.90.174202,PhysRevLett.111.127201,Imbrie_2017, ROS2015420}. 
  We stress that, as occurs for the OTOC~\cite{Swingle_2016,PhysRevA.94.062329,yao2016interferometricapproachprobingfast,PhysRevA.95.012120,Yunger_Halpern_2018,PhysRevE.95.062127,gart,PhysRevX.7.031011}, our findings can be experimentally probed. In fact, our operator-marginal distributions can be accessed directly using standard tools available in current quantum-simulation platforms. The required ingredients are minimal and broadly implemented: the preparation of a doubled system–ancilla state via the Choi--Jamiołkowski isomorphism~\cite{choi1975completely,jamiolkowski1972linear}, controlled forward and backward time evolution under $\hat U$ and $\hat U ^*$, and a measurement protocol based on classical shadows~\cite{huang2020predicting,aaronson2018shadow}. In particular, Bell-pair measurements implemented through local Clifford gates make it possible to extract, from each experimental shot, unbiased estimators of the relevant quantities.
%


The paper is organized as follows: In Sec.~\ref{sec:definitions} we introduce the operator mass and length, and discuss their relations with the nonstabilizerness. In Sec.~\ref{sec:MPS} we explain how to efficiently evaluate them using tensor-network techniques. In Sec.~\ref{sec:results} we evaluate the operator mass and the operator length in the case of the dynamics of an interacting disordered spin chain, finding a logarithmic increase in time when there is strong disorder plus interactions.  We find that the same logarithmic increase in time is displayed by the operator entanglement entropy of an initially localized Hermitian operator. A simple $\ell$-bit model that can be analytically solved and reproduces the logarithmic behavior in time of the operator length is discussed in~\ref{appendix:l-bit}. In Sec.~\ref{sec:experiments}, we discuss the possible experimental implementations of our protocol. In Sec.~\ref{sec:conclusion} we draw our conclusions.

\section{Operator mass and operator length} \label{sec:definitions}

We consider a generic operator $\hat{\mathcal{O}}$ acting on the Hilbert space of $L$ qubits arranged along a one-dimensional geometry, as in the spin-$1/2$ chain analyzed in Sec.~\ref{sec:results}. Any operator can be expanded in the Pauli-string basis $\mathcal{P}_L = \{\hat{Q} = \hat{\sigma}_1^{\alpha_1} \otimes \cdots \otimes \hat{\sigma}_L^{\alpha_L}\}$, where $\alpha_j \in \{0,x,y,z\}$ and $\hat{\sigma}_j^0 \equiv \mathbb{\hat I}$ denotes the identity operator for the $j$th qubit. The expansion coefficients are obtained from the Hilbert--Schmidt inner product
\begin{equation}
  \mathcal{A}_Q = \Tr[\hat{Q}\,\hat{\mathcal{O}}]\,.
  \label{eq:1}
\end{equation}
Using the operator matrix elements $\mathcal{O}_{\{s\},\{s'\}} = \braket{\{s\}|\hat{\mathcal{O}}|\{s'\}}$, Eq.~\eqref{eq:1} reads~\footnote{Notice that after the second equality in Eq.~\eqref{eq:Paulidistr} the sum runs over a single configuration index, because -- given $s_j$ and $\alpha_j$ -- there is one and only one $s_j'(s_j,\alpha_j)$ such that $\braket{s_j'(s_j,\alpha_j)|\hat{\sigma}_j^{\alpha_j}|s_j}\neq 0$~\cite{Russomanno_2025}. This fact is of crucial importance in the exact-diagonalization implementation of the computation of the $\mathcal{A}_Q$, and then of the $m(t)$ and $h(t)$ discussed below (see results in Sec.~\ref{numeris:sec}).}
\begin{equation}
  \mathcal{A}_Q = \sum_{\{s\},\{s'\}} \mathcal{O}_{\{s\},\{s'\}}\,\braket{\{s'\}|\hat{Q}|\{s\}} = \sum_{\{s\}}\mathcal{O}_{\{s\},\{s'\}}\prod_{j=1}^L\braket{s_j'(s_j,\alpha_j)|\hat{\sigma}_j^{\alpha_j}|s_j}\,.
  \label{eq:Paulidistr}
\end{equation}
The Pauli operators form an orthogonal basis under the Hilbert--Schmidt inner product, satisfying
\begin{equation}
  \Tr[\hat{Q} \, \hat{Q}'] = 2^L\,\delta_{\hat{Q},\hat{Q}'}\,.
\end{equation}
Hence, a generic operator admits the expansion
\begin{equation}
  \hat{\mathcal{O}} = \frac{1}{2^L}\sum_{\hat{Q}\in\mathcal{P}_L} \Tr[\hat{\mathcal{O}} \, \hat{Q}]\,\hat{Q}
  = \frac{1}{2^L}\sum_{\hat{Q}\in\mathcal{P}_L} \mathcal{A}_Q\,\hat{Q}\,.
\end{equation}
The coefficients $\{\mathcal{A}_Q\}$ obey the normalization condition
\begin{equation}
  \sum_{\hat{Q}\in\mathcal{P}_L} |\mathcal{A}_Q|^2 = 2^L\,\Tr[\hat{\mathcal{O}}^\dagger\hat{\mathcal{O}}]\,.
\end{equation}
In particular, for a Hermitian operator satisfying $\hat{\mathcal{O}}^2 = \hat{\mathbb{I}}$, one obtains
\begin{equation}
  \sum_{\hat{Q}\in\mathcal{P}_L} |\mathcal{A}_Q|^2 = 2^{2L} = 4^L\,.
\end{equation}

To quantify the delocalization of $\hat{\mathcal{O}}$ in the Pauli basis, one could introduce an operator-space analog of the inverse participation ratio (IPR)~\cite{prosen2007istheEff, Nahum2018PRX}, often referred to as the \emph{operator magic} which is related to the SREs~\cite{PhysRevLett.128.050402}:
\begin{equation}
  \mathcal{M}_2 = -\ln\!\left[
  \frac{\sum_{\hat{Q}\in\mathcal{P}_L} |\mathcal{A}_Q|^4}
  {\left(\sum_{\hat{Q}\in\mathcal{P}_L} |\mathcal{A}_Q|^2\right)^2}
  \right]
  = -\ln\!\left[\frac{\sum_{\hat{Q}\in\mathcal{P}_L} |\mathcal{A}_Q|^4}{4^{2L}}\right]\,.
\end{equation}
This quantity captures how broadly $\hat{\mathcal{O}}$ spreads in the operator basis: $\mathcal{M}_2=0$ for a single Pauli string and increases with the effective number of Pauli components contributing to $\hat{\mathcal{O}}$. Unfortunately, accessing $\mathcal{M}_2$ requires a stochastic sampling~\cite{lami2023Perfect,Haug2023stabilizerentropies,adhikary2025quantumcomplexity}.
To avoid this, we can instead consider complementary measures of operator complexity, namely its \textit{mass} and \textit{length}, both of which can be computed exactly and efficiently within the MPO framework.

\paragraph{Operator mass ---}
Each Pauli string $\hat{Q}$ has a spatial structure, allowing us to assign it a \emph{mass} $m_Q$, defined as the number of non-identity operators in the string:
\begin{equation}
  m_Q = \sum_{j=1}^{L}\big(1 - \delta_{\alpha_j,0}\big)\,.
  \label{eq:operator-mass-definition}
\end{equation}
This definition was introduced in~\cite{PhysRevLett.131.160402} and amounts to count how many non-identity Pauli operators are there in a Pauli string. The average mass of a (time evolving) operator $\mathcal{O}(t)$ is then defined as the value of $m_Q$ weighted by the operator amplitude distribution on the Pauli-string basis, cf.~Eq.~\eqref{eq:Paulidistr}:
\begin{equation}
  m = \frac{\sum_{\hat{Q}\in\mathcal{P}_L} |\mathcal{A}_Q|^2\,m_Q}{\sum_{\hat{Q}\in\mathcal{P}_L} |\mathcal{A}_Q|^2}
  = \frac{1}{4^L}\sum_{\hat{Q}\in\mathcal{P}_L} |\mathcal{A}_Q|^2\,m_Q\,.
\end{equation}

\paragraph{Operator length ---}
Proceeding in a similar manner, to characterize the spatial extent of the operator, we assign to each string $\hat{Q}$ a \emph{length} $h_Q$, defined as the position of its rightmost non-identity operator:
\begin{equation}
  h_Q = \max\{j \in [1,L] : \alpha_j \neq 0\}\,,
  \label{eq:operator-length-definition}
\end{equation}
with $h_Q=0$ if $\hat{Q}=\hat{\mathbb{I}}$. The average operator length is then
\begin{equation}\label{hot:eqn}
  h = \frac{\sum_{\hat{Q}\in\mathcal{P}_L} |\mathcal{A}_Q|^2\,h_Q}{\sum_{\hat{Q}\in\mathcal{P}_L} |\mathcal{A}_Q|^2}
  = \frac{1}{4^L}\sum_{\hat{Q}\in\mathcal{P}_L} |\mathcal{A}_Q|^2\,h_Q\,.
\end{equation}
This quantity probes the spatial spread of operator support along the chain, providing a measure of operator growth and localization in real space.

\textcolor{red}{
An alternative symmetric definition,
$h_Q = \max\{j : \alpha_j \neq 0\} - \min\{j : \alpha_j \neq 0\}$,
can be formulated. However, in our setup the operator is initially
local and its support spreads from a fixed origin under time evolution.
As a consequence, the leftmost non-identity site remains effectively
pinned close to its initial position, while the rightmost site captures
the propagation of the operator front. Therefore, for this specific class of dynamics,
defining the operator length $h(t)$ as the position of the rightmost
non-identity operator provides a direct measure of the spatial extent: the two definitions differ only by an additive constant
and therefore lead to the same scaling behavior. Our choice is thus
sufficient to characterize operator spreading, while being simpler to
evaluate within our framework.
}

\textcolor{red}{
To build intuition, let us consider a simple example of a Pauli-string
decomposition of an operator at time $t$,
\begin{equation}
    \hat{\mathcal{O}} = \sqrt{\frac{5}{10}}\, \hat \sigma^z_2
    + \sqrt{\frac{3}{10}}\, \hat \sigma^x_2 \hat \sigma^x_3
    + \sqrt{\frac{2}{10}}\, \hat \sigma^y_1 \hat \sigma^z_2 \hat \sigma^x_3,
    \label{eq::examplesOp}
\end{equation}
where each term corresponds to a Pauli string with support on different
sites. The squared amplitudes define a probability distribution over
Pauli strings, which induces corresponding distributions for both the
operator length and the operator mass.
For the operator length $h$, one obtains support on sizes $\{2,3\}$ with
probabilities $P(h\!=\!2)=1/2$ and $P(h\!=\!3)=1/2$, reflecting the position of
the rightmost non-identity operator in each string. In contrast, the
operator mass $m$, defined as the number of non-identity sites, has
support on sites $\{1,2,3\}$ with probabilities $P(m\!=\!1)=1/2$, $P(m\!=\!2)=3/10$,
and $P(m\!=\!3)=1/5$.
This example highlights how $h$ and $m$ can probe distinct features
of the operator: $h$ captures its spatial extent, while $m$
characterizes its internal structure, i.e., how many sites are actively
involved. The corresponding expectation values provide a measure of the
typical length and complexity of the operator.
}

\section{MPS formulation for operator spreading}\label{sec:MPS}

The Heisenberg evolution of a local operator typically causes its support to spread throughout the system, a phenomenon that encodes how quantum information and correlations propagate in time. To study this process systematically, it is convenient to represent the operator not as a matrix acting on the physical Hilbert space, but as a vector in an enlarged operator Hilbert space. In this picture, each operator $\hat{\mathcal{O}}$ is mapped to a “ket” $|\hat{\mathcal{O}}\rangle$, and its dynamics is governed by a corresponding Liouvillian (or doubled Hamiltonian) acting on this space. This reformulation enables the application of powerful techniques originally developed for state dynamics, including tensor-network (TN) and entanglement-based methods.

\textcolor{red}{Before proceeding, we clarify the distinction between MPO and MPS in our formulation.
Any physical operator $\hat{\mathcal{O}}$ acting on the Hilbert space is
naturally represented as a MPO. However, by vectorizing the operator in
the enlarged (Liouville) Hilbert space, $\hat{\mathcal{O}}$ can be mapped
to a state $|\mathcal{\hat O}\rangle$, which admits an MPS representation. In this work, we use the MPO language when referring to the physical
operator, and the MPS language when discussing its representation as a
state in the enlarged space and the associated numerical computations.
This mapping allows us to exploit standard MPS techniques to efficiently
simulate operator dynamics.}

Within this framework, matrix product states (MPS) provide an efficient way to encode and simulate the exponentially large operator space arising in many-body systems~\cite{collura2024tensor}. The MPS formalism allows one to represent local operators and their time evolution compactly, capturing the buildup of correlations and operator entanglement with a tunable bond dimension. This representation is particularly advantageous to study the operator growth, as it provides both a conceptual mapping between operators and states and a practical computational tool for tracking delocalization, scrambling, and localization dynamics.
In this section, we reformulate the quantities introduced in Sec.~\ref{sec:definitions} (characterizing spatial delocalization of an operator, under time evolution) within the MPS framework.
Our approach is particularly convenient for numerical simulations of many-body dynamics, as it enables an efficient representation and manipulation of operators acting on large Hilbert spaces.

\paragraph{Normalized Pauli basis \& Pauli tensor ---}
To remove the redundant normalization factor $2^L$ that appears in the definition of probabilities in the standard Pauli basis, we adopt a \emph{normalized Pauli basis}.
For a spin-1/2 system, we define the normalized single-site operators as
\begin{equation}
    \hat{\varsigma}^{\mu} = \frac{\hat{\sigma}^{\mu}}{\sqrt{2}},
    \qquad \mu \in \{0, x, y, z\} \,.
\end{equation}
The so-called \emph{Pauli tensor} naturally emerges from this set and can be represented diagrammatically as
\begin{equation*}
\begin{tikzpicture}[baseline=-0.6ex,node distance=0.8cm, scale=0.75]
    \node[tensorCir, fill=pastelRed!150, minimum size=0.8cm] (A) {$\varsigma$};
    \draw[index] (A) -- ++(0,1) node[above] {$i$};
    \draw[index] (A) -- ++(0,-1) node[below] {$j$};
    \draw[index] (A) -- ++(1,0) node[right] {$\mu$};
\end{tikzpicture} 
=\frac{1}{\sqrt{2}} \Biggl\{
\begin{tikzpicture}[baseline=-0.6ex, scale=0.75]
    \node[tensorCir, fill=pastelCoral!10, minimum size=0.8cm] (I) {$\mathbb{I}$};
    \draw[index] (I) -- ++(0,1) node[above]{$i$};
    \draw[index] (I) -- ++(0,-1) node[below]{$j$};
\end{tikzpicture}\hspace{2mm},\hspace{2mm}
\begin{tikzpicture}[baseline=-0.6ex, scale=0.75]
    \node[tensorCir, fill=pastelBlue!150, minimum size=0.8cm] (X) {$\mathbb{X}$};
    \draw[index] (X) -- ++(0,1) node[above]{$i$};
    \draw[index] (X) -- ++(0,-1) node[below]{$j$};
\end{tikzpicture}\hspace{2mm},\hspace{2mm}
\begin{tikzpicture}[baseline=-0.6ex, scale=0.75]
    \node[tensorCir, fill=tensorGreen!125, minimum size=0.8cm] (Y){$\mathbb{Y}$};
    \draw[index] (Y) -- ++(0,1) node[above]{$i$};
    \draw[index] (Y) -- ++(0,-1) node[below]{$j$};
\end{tikzpicture}\hspace{2mm},\hspace{2mm}
\begin{tikzpicture}[baseline=-0.6ex, scale=0.75]
    \node[tensorCir, fill=tensorOrange!90, minimum size=0.8cm] (Z){$\mathbb{Z}$};
    \draw[index] (Z) -- ++(0,1) node[above]{$i$};
    \draw[index] (Z) -- ++(0,-1) node[below]{$j$};
\end{tikzpicture}
\Biggr\}.
\end{equation*}
The Pauli tensor satisfies the orthogonality and completeness relations,
which can be represented graphically as~\footnote{The first equality below in formulae reads $\Tr[\varsigma^\mu\varsigma^\nu]=\delta^{\mu\nu}$, the second $\sum_{\mu}\varsigma_{ij}^\nu\varsigma_{i'j'}^\mu = \delta_{ij}\delta_{i'j'}$. These equalities are not difficult to check using the definitions of the Pauli matrices.}
\begin{center}
\begin{tikzpicture}[baseline=-0.6ex, node distance=0.4cm, scale=0.75]
    \node[tensorCir, fill=pastelRed!150, minimum size=0.8cm](At){$\varsigma^{*}$};
    \draw[index] (At) -- ++(0,1);
    \draw[index] (At) -- ++(0,-1);
    \draw[index] (At) -- ++(-1,0) node[left]{$\mu$};
    \node[tensorCir, fill=pastelRed!150, minimum size=0.8cm,right=of At](A){$\varsigma$};
    \draw[index] (A) -- ++(0,1);
    \draw[index] (A) -- ++(0,-1);
    \draw[index] (A) -- ++(1,0) node[right]{$\nu$};
    \draw[index] ($(At)+(0,1)$) -- ($(A)+(0,1)$);
    \draw[index] ($(At)+(0,-1)$) -- ($(A)+(0,-1)$);
\end{tikzpicture}
$=$
\begin{tikzpicture}[baseline=-0.6ex]
    \draw node[left]{$\mu$}[index] (0,0) -- (1.5,0) node[right]{$\nu$};
\end{tikzpicture}
\hspace{2mm},\hspace{2mm}
\begin{tikzpicture}[baseline=-0.6ex, node distance=0.4cm, scale=0.75]
    \node[tensorCir, fill=pastelRed!150, minimum size=0.8cm](A){$\varsigma$};
    \draw[index] (A) -- ++(0,1) node[above]{$i$};
    \draw[index] (A) -- ++(0,-1) node[below]{$j$};
    \node[tensorCir, fill=pastelRed!150, minimum size=0.8cm, right=of A] (At){$\varsigma^{*}$};
    \draw[index] (At) -- ++(0,1) node[above]{$i'$};
    \draw[index] (At) -- ++(0,-1) node[below]{$j'$};
    \draw[index] (A) -- (At);
\end{tikzpicture}
\hspace{2mm}$=$
\begin{tikzpicture}[baseline=-0.6ex, scale=0.75]
    \draw[index] (-1,1) .. controls (-0.5,0) and (.5,0) .. (1,1);
    \draw[index] (-1,-1) .. controls (-0.5,0) and (.5,0) .. (1,-1);
    \node[above left] at (-1,1) {$i$};
    \node[above right] at (1,1) {$i'$};
    \node[below left] at (-1,-1) {$j$};
    \node[below right] at (1,-1) {$j'$};
\end{tikzpicture}
\end{center}
As discussed in the previous section, any generic single-site operator $\hat{\mathcal{O}}=\sum_{i,j} O_{ij}|i\rangle\langle j|$ can be expressed in the normalized Pauli-string basis as
\[
\hat{\mathcal{O}} = \sum_{\mu} A_{\mu}\,\hat{\varsigma}^{\mu},
\qquad
A_{\mu} = \mathrm{Tr}\big[ \hat{\mathcal{O}}\,\hat{\varsigma}^{\mu}\big].
\]
This representation can be depicted diagrammatically as the contraction between the Pauli tensor and the tensor representation of $\hat{\mathcal{O}}$:
\begin{equation*}
\begin{tikzpicture}[baseline=-0.6ex, node distance=1cm, scale=0.75]
    \node[tensorCir, fill=tensorTeal!80, minimum size=0.8cm] (V){$A$};
    \draw[index] (V) -- ++(0,-1) node[below]{$\mu$};
\end{tikzpicture}
\hspace{1.5mm} = \hspace{1.5mm}
\begin{tikzpicture}[baseline=-0.6ex, node distance=1cm, scale=0.75]
    \node[tensorCir, fill=pastelRed!150, minimum size=0.8cm] (A){$\varsigma^*$};
    \draw[index] (A) -- ++(0,1);
    \draw[index] (A) -- ++(0,-1);
    \draw[index] (A) -- ++(1,0) node[right]{$\mu$};
    \node[tensorCir, fill=pastelCyan!100, minimum size=0.8cm, right=of A] (Z){$O$};
    \draw[index] (Z) -- ++(0,1); 
    \draw[index] (Z) -- ++(0,-1);
    \draw[index] ($(A)+(0,1)$) -- ($(Z)+(0,1)$);
    \draw[index] ($(A)+(0,-1)$) -- ($(Z)+(0,-1)$);
\end{tikzpicture}.
\end{equation*}
The generalization to a chain of $L$ sites follows straightforwardly by introducing the set of normalized Pauli strings
\[
\hat{P}_{\boldsymbol{\mu}} = \bigotimes_{j=1}^{L}\hat{\varsigma}^{\mu_j},
\]
satisfying the orthonormality condition
$\mathrm{Tr}\!\big[\hat{P}_{\boldsymbol{\mu}}\hat{P}_{\boldsymbol{\nu}}\big] = \delta_{\boldsymbol{\mu}\boldsymbol{\nu}}$.
Any operator acting on the full Hilbert space can thus be expanded as~\footnote{The $A_{\boldsymbol{\mu}}$ defined in Eq.~\eqref{Amu:eqn} are related to the $\mathcal{A}_Q$ defined in Eq.~\eqref{eq:1} by the relation $A_{\boldsymbol{\mu}} = \mathcal{A}_{Q_{\boldsymbol{\mu}}} / 2^{L/2}$, with $Q_{\boldsymbol{\mu}}\equiv2^{L/2}P_{\boldsymbol{\mu}}$.}
\begin{equation}\label{Amu:eqn}
\hat{\mathcal{O}} = \sum_{\boldsymbol{\mu}} A_{\boldsymbol{\mu}}\, \hat{P}_{\boldsymbol{\mu}},
\qquad
A_{\boldsymbol{\mu}} = \mathrm{Tr} \big[\hat{\mathcal{O}}\,\hat{P}_{\boldsymbol{\mu}} \big].
\end{equation}
For a Hermitian operator $\hat{\mathcal{O}}$, the coefficients $A_{\boldsymbol{\mu}}$ are real.  
If we further impose the normalization condition $\mathrm{Tr} [\hat{\mathcal{O}}^2] = 1$, these coefficients satisfy $\sum_{\boldsymbol{\mu}} (A_{\boldsymbol{\mu}})^2 = 1$.
Hence, $\hat{\mathcal{O}}$ can be mapped to a normalized state vector
\begin{equation}\label{statop:eqn}
|\hat{\mathcal{O}}\rangle = \sum_{\boldsymbol{\mu}} A_{\boldsymbol{\mu}} |\boldsymbol{\mu}\rangle
\end{equation}
in an auxiliary Hilbert space of dimension $4^L$. This mapping, often referred to as the \emph{Pauli–Liouville} or \emph{Choi} representation, allows one to apply tensor-network techniques developed for pure states to the study of the operator dynamics. In practice, the $4^L$-component coefficient tensor $A_{\boldsymbol{\mu}}$ is not stored explicitly, but represented in a compressed tensor-network form, as discussed below.

\subsection{MPS representation of operators}
The coefficient tensor $A_{\boldsymbol{\mu}}$ can be factorized in a MPS form with bond dimension $\chi$,
\begin{equation}
    A_{\mu_1 \ldots \mu_L}
    = \mathbf{v}_L^{\mathsf{T}} \, 
      \Bigg( \prod_{j=1}^{L} \mathbf{A}^{[\mu_j]}_j \Bigg)
      \, \mathbf{v}_R,
\end{equation}
where $\mathbf{A}^{[\mu_j]}_j$ are rank-3 tensors associated with each site
and $\mathbf{v}_{L,R}$ are boundary vectors.
The bond dimension $\chi$ of this operator state reflects its internal tensor structure: for instance, an operator that factorizes as a product of local terms has $\chi=1$.
In this language, the operator $\hat{\mathcal{O}}$ is treated as an MPS in the \emph{operator Hilbert space}, and its components $A_{\mu_1\ldots\mu_L}$ are interpreted as the amplitudes of a normalized quantum state. Equivalently, the operator-state $|\hat O\rangle \in (\mathbb C^{4})^{\otimes L}$ admits an MPS representation in the auxiliary (operator) Hilbert space, given by
\[
|\hat{\mathcal{O}}\rangle = 
\sum_{\mu_1, \ldots, \mu_L}\hspace{2mm}
\begin{tikzpicture}[baseline=-0.6ex,node distance=0.8cm, scale=.75]
\centering
  \node[tensorCir, fill=tensorTeal!80, minimum size=.8cm] (A) {};
  \node[tensorCir, fill=tensorTeal!80, minimum size=.8cm, right=of A] (B) {};
  \node[tensorCir, fill=tensorTeal!80, minimum size=.8cm, right=of B] (C) {};
  \node[draw=none, right=of C](D){};
  \node[tensorCir, fill=tensorTeal!80, minimum size=.8cm, right=of D] (E) {};
  \draw[index] (A) -- node[below] {} (B);
  \draw[index] (B) -- node[above] {$\chi$} (C);
  \draw[index] (C) -- ++(1,0);
  \draw[index] (E) -- ++(-1,0);
  \draw[index] (A) -- ++(0,-1) node[below]{$\mu_{1}$};
  \draw[index] (B) -- ++(0,-1) node[below]{$\mu_{2}$};
  \draw[index] (C) -- ++(0,-1) node[below]{};
  \draw[index] (E) -- ++(0,-1) node[below]{$\mu_{L}$};
  \draw[index, densely dashed] ($(C)+(1.25,0)$) -- ($(E)+(-1.25,0)$);
\end{tikzpicture}
\hspace{3mm}
|\mu_{1}\mu_{2}\cdot\cdot\cdot\mu_{L}\rangle
\]

We assume the normalization condition $\Tr[\hat{\mathcal{O}}^2]=1$, which in tensor network notation
reads:
\begin{equation*}
    \Tr[\hat{\mathcal{O}}^{2}] \hspace{1mm} =\hspace{1mm}
\begin{tikzpicture}[baseline=-5.0ex,node distance=0.8cm, scale=.75]
\centering
  \node[tensorCir, fill=tensorTeal!80] (A) {};
  \node[tensorCir, fill=tensorTeal!80, below=of A] (At) {};
  \node[tensorCir, fill=tensorTeal!80, right=of A] (B) {};
  \node[tensorCir, fill=tensorTeal!80, below=of B] (Bt) {};
  \node[tensorCir, fill=tensorTeal!80, right=of C] (D) {};
  \node[tensorCir, fill=tensorTeal!80, below=of D] (Dt) {};

  \draw[index] (A) -- (B);
  \draw[index] (At) -- (Bt);
  \draw[index] (B) -- ++(1,0);
  \draw[index] (Bt) -- ++(1,0);
  
  \draw[index] (A) -- (At);
  \draw[index] (B) -- (Bt);
  \draw[index, dashed] ($(B)+(1.5,0)$) -- ($(D)+(-1.5,0)$);
  \draw[index, dashed] ($(Bt)+(1.5,0)$) -- ($(Dt)+(-1.5,0)$);
  \draw[index] (D) -- ++(-1,0);
  \draw[index] (Dt) -- ++(-1,0);
  \draw[index] (D) -- (Dt);
\end{tikzpicture}
\hspace{1mm} =\hspace{1mm} 1,
\end{equation*}
Under this normalization, the squared amplitudes $A_{\boldsymbol{\mu}}^2$ acquire a probabilistic interpretation:
\begin{equation}
    \Pi_{\hat{\mathcal{O}}}(\boldsymbol{\mu})
    = A_{\boldsymbol{\mu}}^2
    = \big( \Tr\big[ \hat{\mathcal{O}} \, \hat{P}_{\boldsymbol{\mu}} \big] \big)^2
    \ge 0, \qquad
    \sum_{\boldsymbol{\mu}} \Pi_{\hat{\mathcal{O}}}(\boldsymbol{\mu}) = 1.
\end{equation}
Hence, the MPS coefficients define a probability distribution over the operator indices $\boldsymbol{\mu}$,
providing a direct probabilistic interpretation of the operator’s MPS wave function.

A generic amplitude can be computed diagrammatically as
\begin{equation*}
A_{\boldsymbol{\mu}}=\Tr[\hat{\mathcal{O}}\hat{\varsigma}_{1}^{\mu_{1}}\ldots\hat{\varsigma}_{L}^{\mu_{L}}] \hspace{1mm} = \hspace{1mm} 
\begin{tikzpicture}[baseline=-0.6ex, node distance=0.8cm, scale=0.75]
    \node[tensorCir, fill=tensorTeal!80] (A) {};
    \node[tensorCir, fill=tensorTeal!80, right=of A] (B) {};
    \node[tensorCir, fill=tensorTeal!80, right=of C] (D){};

    \draw[index] (A) -- (B);
    \draw[index] (B) -- ++(1,0);
    \draw[index] (D) -- ++(-1,0);

    \draw[index] (A) -- ++(0,-1) node[below]{$\mu_{1}$};
    \draw[index] (B) -- ++(0,-1) node[below]{$\mu_{2}$};
    \draw[index] (D) -- ++(0,-1) node[below]{$\mu_{L}$};

    \draw[index, dashed] ($(B)+(1.5,0)$) -- ($(D)+(-1.5,0)$);
\end{tikzpicture} \,.
\end{equation*}
This probabilistic interpretation is useful for defining observables that characterize the spatial extent of the operator $\hat{\mathcal{O}}$—such as its \emph{length} and \emph{mass}—within the tensor-network formalism introduced below.
%
Each coefficient $A_{\boldsymbol{\mu}}(t)$ in the expansion of $\hat{\mathcal{O}}$can be viewed as the amplitude of a normalized wave function in the so-called operator Hilbert space, allowing the growth of operators to be tracked through the statistics of these coefficients \cite{Nahum2018PRX,vonKeyserlingk2018PRL}. $A_{\boldsymbol{\mu}}(t)$ are furthermore the key for evaluating the operator mass and length and so are very important in the study of operator delocalization in quantum dynamics, where the increase of operator length provides information on scrambling. In the subsections below we are going to explain how to use the $A_{\boldsymbol{\mu}}(t)$ in an MPS framework to evaluate the operator length (Sec.~\ref{olen:sec}), the operator mass (Sec.~\ref{omas:sec}) and the operator entanglement entropy (Sec.~\ref{oen:sec}).

\subsection{Operator length}\label{olen:sec}
As defined in Eq.~\eqref{eq:operator-length-definition}, each Pauli string $\hat{P}_{\boldsymbol{\mu}}$ can be assigned a ``length''
$h_{\boldsymbol{\mu}}$, defined as the largest site index $j$ such that
$\mu_j \neq 0$, with $h_{\boldsymbol{\mu}} = 0$ if all $\mu_j = 0$.
The average operator length is then obtained as the expectation value
\begin{equation}
    h_{\hat{\mathcal{O}}}
    = \sum_{\boldsymbol{\mu}} h_{\boldsymbol{\mu}} \, \Pi_{\hat{\mathcal{O}}}(\boldsymbol{\mu})
    = \sum_{l=0}^{L} l \, P_{\hat{\mathcal{O}}}(l),
    \label{eq:operator-length-average}
\end{equation}
where
\begin{equation}
    P_{\hat{\mathcal{O}}}(l)
    = \sum_{\boldsymbol{\mu}} \delta_{h_{\boldsymbol{\mu}},\,l} \,
      \Pi_{\hat{\mathcal{O}}}(\boldsymbol{\mu})
\end{equation}
is the probability that $\hat{\mathcal{O}}$ has support up to site $l$. This probability is easy to compute in our tensor network-formalism, indeed one has

\begin{equation*}
\begin{split}
P_{\hat{\mathcal{O}}}(l) & = \begin{tikzpicture}[baseline=-8.0ex, node distance=0.4cm, scale=0.55]
    \node[tensorCir, fill=tensorTeal!80, minimum size=7mm] (A) {}; 
    \node[above=2pt of A] {$1$};
    \node[draw=none, right=of A] (B) {};
    \node[tensorCir, fill=tensorTeal!80, minimum size=7mm, right=of B] (C) {};
    \node[above=2pt of C] {$l$};
    \node[tensorCir, fill=tensorTeal!80, minimum size=7mm, right=of C] (D) {};
    \node[above=2pt of D] {$l+1$};
    \node[draw=none, right=of D] (E) {};
    \node[tensorCir, fill=tensorTeal!80, minimum size=7mm, right=of E] (F) {};
    \node[above=2pt of F] {$L$};
    \node[tensorCir, fill=tensorTeal!80, minimum size=7mm, below=1.4cm of A] (At) {};
    \node[tensorCir, fill=tensorTeal!80, minimum size=7mm, below=1.4cm of C] (Ct) {};
    \node[tensorCir, fill=tensorTeal!80, minimum size=7mm, below=1.4cm of D] (Dt) {};
    \node[tensorCir, fill=tensorTeal!80, minimum size=7mm, below=1.4cm of F] (Ft) {};
    \draw[index] (A) -- ++(1,0);
    \draw[index] (C) -- ++(-1,0);
    \draw[index] (C) -- (D);
    \draw[index] (D) -- ++(1,0);
    \draw[index] (F) -- ++(-1,0);
    \draw[index] (At) -- ++(1,0);
    \draw[index] (Ct) -- ++(-1,0);
    \draw[index] (Ct) -- (Dt);
    \draw[index] (Dt) -- ++(1,0);
    \draw[index] (Ft) -- ++(-1,0);
    \draw[index] (A) -- (At);
    \draw[index] (C) -- (Ct);
    \draw[index] (D) -- ++(0,-1.5);
    \draw[index] (Dt) -- ++(0,1.5);
    \draw[index] (F) -- ++(0,-1.5);
    \draw[index] (Ft) -- ++(0,1.5);
    \node[tensorTri, fill=pastelMarron!100, draw=pastelMarron!100, text=white, below=.6cm of D, minimum size=0.1mm, shape border rotate=-180] (Tru1) {$0$};
    \node[tensorTri, fill=pastelMarron!100, draw=pastelMarron!100, text=white, above=.6cm of Dt, minimum size=0.1mm] (Trd1) {$0$};
    \node[tensorTri, fill=pastelMarron!100, draw=pastelMarron!100, text=white, below=.6cm of F, minimum size=0.1mm, shape border rotate=-180] (Tru2) {$0$};
    \node[tensorTri, fill=pastelMarron!100, draw=pastelMarron!100, text=white, above=.6cm of Ft, minimum size=0.1mm] (Trd2) {$0$};
    \draw[index, dashed] ($(A)+(1.4,0)$) -- ($(C)+(-1.4,0)$);
    \draw[index, dashed] ($(D)+(1.4,0)$) -- ($(F)+(-1.4,0)$);
    \draw[index, dashed] ($(At)+(1.4,0)$) -- ($(Ct)+(-1.4,0)$);
    \draw[index, dashed] ($(Dt)+(1.4,0)$) -- ($(Ft)+(-1.4,0)$);
    \draw[decorate, decoration={brace, amplitude=5pt, mirror}, very thick]($(At)+(-0.5,-1)$) -- ($(Ft)+(0.5,-1)$)
    node[midway, below=4pt] {$\tilde{S}_{2}(l)$};
\end{tikzpicture} \hspace{1mm}-\hspace{1mm}
\begin{tikzpicture}[baseline=-8.0ex, node distance=0.4cm, scale=0.55]
    \node[tensorCir, fill=tensorTeal!80, minimum size=7mm] (A) {}; 
    \node[above=2pt of A] {$1$};
    \node[draw=none, right=of A] (B) {};
    \node[tensorCir, fill=tensorTeal!80, minimum size=7mm, right=of B] (C) {};
    \node[above=2pt of C] {$l-1$};
    \node[tensorCir, fill=tensorTeal!80, minimum size=7mm, right=of C] (D) {};
    \node[above=2pt of D] {$l$};
    \node[draw=none, right=of D] (E) {};
    \node[tensorCir, fill=tensorTeal!80, minimum size=7mm, right=of E] (F) {};
    \node[above=2pt of F] {$L$};
    \node[tensorCir, fill=tensorTeal!80, minimum size=7mm, below=1.4cm of A] (At) {};
    \node[tensorCir, fill=tensorTeal!80, minimum size=7mm, below=1.4cm of C] (Ct) {};
    \node[tensorCir, fill=tensorTeal!80, minimum size=7mm, below=1.4cm of D] (Dt) {};
    \node[tensorCir, fill=tensorTeal!80, minimum size=7mm, below=1.4cm of F] (Ft) {};
    \draw[index] (A) -- ++(1,0);
    \draw[index] (C) -- ++(-1,0);
    \draw[index] (C) -- (D);
    \draw[index] (D) -- ++(1,0);
    \draw[index] (F) -- ++(-1,0);
    \draw[index] (At) -- ++(1,0);
    \draw[index] (Ct) -- ++(-1,0);
    \draw[index] (Ct) -- (Dt);
    \draw[index] (Dt) -- ++(1,0);
    \draw[index] (Ft) -- ++(-1,0);
    \draw[index] (A) -- (At);
    \draw[index] (C) -- (Ct);
    \draw[index] (D) -- ++(0,-1.5);
    \draw[index] (Dt) -- ++(0,1.5);
    \draw[index] (F) -- ++(0,-1.5);
    \draw[index] (Ft) -- ++(0,1.5);
    \node[tensorTri, fill=pastelMarron!100, draw=pastelMarron!100, text=white, below=.6cm of D, minimum size=0.1mm, shape border rotate=-180] (Tru1) {$0$};
    \node[tensorTri, fill=pastelMarron!100, draw=pastelMarron!100, text=white, above=.6cm of Dt, minimum size=0.1mm] (Trd1) {$0$};
    \node[tensorTri, fill=pastelMarron!100, draw=pastelMarron!100, text=white, below=.6cm of F, minimum size=0.1mm, shape border rotate=-180] (Tru2) {$0$};
    \node[tensorTri, fill=pastelMarron!100, draw=pastelMarron!100, text=white, above=.6cm of Ft, minimum size=.1mm] (Trd2) {$0$};
    \draw[index, dashed] ($(A)+(1.4,0)$) -- ($(C)+(-1.4,0)$);
    \draw[index, dashed] ($(D)+(1.4,0)$) -- ($(F)+(-1.4,0)$);
    \draw[index, dashed] ($(At)+(1.4,0)$) -- ($(Ct)+(-1.4,0)$);
    \draw[index, dashed] ($(Dt)+(1.4,0)$) -- ($(Ft)+(-1.4,0)$);
    \draw[decorate, decoration={brace, amplitude=5pt, mirror}, very thick]($(At)+(-0.5,-1)$) -- ($(Ft)+(0.5,-1)$)
    node[midway, below=4pt] {$\tilde{S}_{2}(l-1)$};
\end{tikzpicture} \\
& = \hspace{5mm}\begin{tikzpicture}[baseline=-8.0ex, node distance=1.25cm, scale=0.6]
    \node[tensorTri, fill=tensorTeal!80, minimum size=15mm,shape border rotate=-90] (A) {}; 
    \node[above=2pt of A] {$l+1$};
    \node[tensorTri, fill=tensorTeal!80, minimum size=15mm, shape border rotate=-90, right=of B] (C) {};
    \node[above=2pt of C] {$L$};
    \node[tensorTri, fill=tensorTeal!80, minimum size=15mm, shape border rotate=-90, below=2.8cm of A] (At) {};
    \node[tensorTri, fill=tensorTeal!80, minimum size=15mm, shape border rotate=-90, below=2.8cm of C] (Ct) {};
    \draw[index] (A) -- ++(1.25,0);
    \draw[index] (C) -- ++(-1,0);
    \draw[index] (At) -- ++(1.25,0);
    \draw[index] (Ct) -- ++(-1,0);
    \node[tensorTri, fill=pastelMarron!100, draw=pastelMarron!100, text=white, below=0.75cm of A, minimum size=0.1mm, shape border rotate=-180] (Tru1) {$0$};
    \node[tensorTri, fill=pastelMarron!100, draw=pastelMarron!100, text=white, above=0.75cm of At, minimum size=0.1mm] (Trd1) {$0$};
    \node[tensorTri, fill=pastelMarron!100, draw=pastelMarron!100, text=white, below=0.75cm of C, minimum size=0.1mm, shape border rotate=-180] (Tru2) {$0$};
    \node[tensorTri, fill=pastelMarron!100, draw=pastelMarron!100, text=white, above=0.75cm of Ct, minimum size=0.1mm] (Trd2) {$0$};
    \draw[index] (A) -- (Tru1);
    \draw[index] (At) -- (Trd1);
    \draw[index] (C) -- (Tru2);
    \draw[index] (Ct) -- (Trd2);
    \draw[index, dashed] ($(A)+(1.45,0)$) -- ($(C)+(-1.2,0)$);
    \draw[index, dashed] ($(At)+(1.45,0)$) -- ($(Ct)+(-1.2,0)$);
    \draw[index] (A) -- ++(-1.25,0);
    \draw[index] (At) -- ++(-1.25,0);
    \draw[index] ($(A)+(-1.25,0)$) -- ($(At)+(-1.25,0)$);
    \draw[decorate, decoration={brace, amplitude=5pt, mirror}, very thick]($(At)+(-1.5,-1)$) -- ($(Ct)+(1,-1)$)
    node[midway, below=4pt] {$\tilde{S}_{2}(l)$};
\end{tikzpicture} \hspace{5mm}-\hspace{7mm}
\begin{tikzpicture}[baseline=-8.0ex, node distance=1.25cm, scale=0.6]
    \node[tensorTri, fill=tensorTeal!80, minimum size=15mm,shape border rotate=-90] (A) {}; 
    \node[above=2pt of A] {$l$};
    \node[tensorTri, fill=tensorTeal!80, minimum size=15mm, shape border rotate=-90, right=of B] (C) {};
    \node[above=2pt of C] {$L$};
    \node[tensorTri, fill=tensorTeal!80, minimum size=15mm, shape border rotate=-90, below=2.8cm of A] (At) {};
    \node[tensorTri, fill=tensorTeal!80, minimum size=15mm, shape border rotate=-90, below=2.8cm of C] (Ct) {};
    \draw[index] (A) -- ++(1.25,0);
    \draw[index] (C) -- ++(-1,0);
    \draw[index] (At) -- ++(1.25,0);
    \draw[index] (Ct) -- ++(-1,0);
    \node[tensorTri, fill=pastelMarron!100, draw=pastelMarron!100, text=white, below=.75cm of A, minimum size=0.1mm, shape border rotate=-180] (Tru1) {$0$};
    \node[tensorTri, fill=pastelMarron!100, draw=pastelMarron!100, text=white, above=.75cm of At, minimum size=0.1mm] (Trd1) {$0$};
    \node[tensorTri, fill=pastelMarron!100, draw=pastelMarron!100, text=white, below=.75cm of C, minimum size=0.1mm, shape border rotate=-180] (Tru2) {$0$};
    \node[tensorTri, fill=pastelMarron!100, draw=pastelMarron!100, text=white, above=.75cm of Ct, minimum size=0.1mm] (Trd2) {$0$};
    \draw[index] (A) -- (Tru1);
    \draw[index] (At) -- (Trd1);
    \draw[index] (C) -- (Tru2);
    \draw[index] (Ct) -- (Trd2);
    \draw[index, dashed] ($(A)+(1.45,0)$) -- ($(C)+(-1.2,0)$);
    \draw[index, dashed] ($(At)+(1.45,0)$) -- ($(Ct)+(-1.2,0)$);
    \draw[index] (A) -- ++(-1.25,0);
    \draw[index] (At) -- ++(-1.25,0);
    \draw[index] ($(A)+(-1.25,0)$) -- ($(At)+(-1.25,0)$);
    \draw[decorate, decoration={brace, amplitude=5pt, mirror}, very thick]($(At)+(-1.5,-1)$) -- ($(Ct)+(1,-1)$)
    node[midway, below=4pt] {$\tilde{S}_{2}(l-1)$};
\end{tikzpicture} \,.
\end{split}
\end{equation*}
When $\hat{\mathcal{O}}$ is represented as a MPS in the normalized Pauli basis,
the quantities $P_{\hat{\mathcal{O}}}(l)$ can be efficiently computed from the
unnormalized Rényi-2 entropies
\begin{equation}
    \tilde{S}_2(l)
    = \mathrm{Tr}\!\left[ \hat{\mathcal{O}}_{|l}^2 \right],
    \qquad
    \hat{\mathcal{O}}_{|l} = \frac{\mathrm{Tr}_{j>l}(\hat{\mathcal{O}})}{2^{(L-l)/2}},
\end{equation}
by exploiting the recurrence relation
\begin{equation}
    P_{\hat{\mathcal{O}}}(l)
    = \tilde{S}_2(l) - \tilde{S}_2(l-1),
    \qquad
    \tilde{S}_2(-1) \equiv 0.
\end{equation}
This allows the average operator length to be written compactly as
\begin{equation}
    h_{\hat{\mathcal{O}}} = L - \sum_{l=0}^{L-1} \tilde{S}_2(l),
    \label{eq:length-vs-renyi}
\end{equation}
where $\tilde{S}_2(L) = \mathrm{Tr}(\hat{\mathcal{O}}^2) = 1$ simply enforces normalization.
The use of the left-canonical form of the MPS reduces the numerical cost of evaluating Eq.~\eqref{eq:length-vs-renyi}
from $\mathcal{O}(L\chi^3)$ to $\mathcal{O}(L\chi^2)$.

The quantity $h_{\hat{\mathcal{O}}}$ thus provides an efficient and physically transparent measure of the spatial support of the operator in the Pauli-string basis.
%
\subsection{Operator mass}\label{omas:sec}
As introduced in Eq.~\eqref{eq:operator-mass-definition}, we define the \emph{mass} of a Pauli string $\hat{P}_{\boldsymbol{\mu}}$
as the number of sites on which the operator acts nontrivially, simply counting how many local Pauli matrices different from the identity appear in the string.
The average operator mass is obtained as
\begin{equation}
    m_{\hat{\mathcal{O}}}
    = \sum_{\boldsymbol{\mu}} m_{\boldsymbol{\mu}} \, \Pi_{\hat{\mathcal{O}}}(\boldsymbol{\mu})
    = \sum_{m=0}^{L} m \, P_{\hat{\mathcal{O}}}(m),
    \label{eq:operator-mass-average}
\end{equation}
where
\begin{equation}
    P_{\hat{\mathcal{O}}}(m)
    = \sum_{\boldsymbol{\mu}}
      \delta_{m_{\boldsymbol{\mu}},\,m} \,
      \Pi_{\hat{\mathcal{O}}}(\boldsymbol{\mu})
\end{equation}
is the probability that $\hat{\mathcal{O}}$ contains exactly $m$ nontrivial local operators.

In the operator Hilbert space, the total mass operator can be expressed as a diagonal
superoperator acting on the normalized Pauli strings:
\begin{equation}
    \hat{M}
    = \sum_{\boldsymbol{\mu}}
      m_{\boldsymbol{\mu}} \, |\boldsymbol{\mu}\rangle \langle \boldsymbol{\mu}|
    = \sum_{j=1}^{L} \big( \, \hat{\mathbb{I}}_j - |0\rangle\langle 0|_j \big),
    \label{eq:mass-superoperator}
\end{equation}
where $|\boldsymbol{\mu}\rangle$ denotes the vectorized Pauli string $\hat{P}_{\boldsymbol{\mu}}$.
The expectation value of $\hat{M}$ over the operator state $|\hat{\mathcal{O}}\rangle$
yields the average mass,
\begin{equation}
    m_{\hat{\mathcal{O}}} = \langle \hat{\mathcal{O}} | \hat{M} | \hat{\mathcal{O}} \rangle.
\end{equation}

Since $\hat{M}$ is diagonal in the Pauli basis and factorizes over sites, it can be written as a simple MPO
with local tensors of auxiliary dimension $2$.
This structure enables an efficient tensor-network contraction
for evaluating $m_{\hat{\mathcal{O}}}$ with cost $\mathcal{O}(L\chi^3)$.

\begin{equation*}
\hat{M} = 
\begin{tikzpicture}[baseline=-0.6ex, node distance=1cm, scale=0.75]
    \node[tensorSqua, fill=pastelYellow!90] (A) {};
    \node[tensorSqua, fill=pastelYellow!90, right=of A] (B) {};
    \node[draw=none, right=of B] (C) {};
    \node[tensorSqua, fill=pastelYellow!90, right=of C] (D) {};

    \draw[index] (A) -- (B);
    \draw[index] (B) -- ++(1,0);
    \draw[index, dashed] ($(B)+(1.25,0)$) -- ($(D)+(-1.25,0)$);
    \draw[index] (D) -- ++(-1,0);

    \draw[index] (A) -- ++(0,1);
    \draw[index] (B) -- ++(0,1);
    \draw[index] (D) -- ++(0,1);

    \draw[index] ($(A)+(0,1)$) -- ++($(-1.25,0)$);
    \draw[index] ($(B)+(0,1)$) -- ++($(-1.25,0)$);
    \draw[index] ($(D)+(0,1)$) -- ++($(-1.25,0)$);

    \draw[index] ($(A)+(-1.25,1.5)$) node[above]{$\nu_{1}$} -- ++(0,-3) node[below]{$\mu_{1}$};
    \draw[index] ($(B)+(-1.25,1.5)$) node[above]{$\nu_{2}$} -- ++(0,-3) node[below]{$\mu_{2}$};
    \draw[index] ($(D)+(-1.25,1.5)$) node[above]{$\nu_{L}$} -- ++(0,-3) node[below]{$\mu_{L}$};

    \filldraw[black] ($(A)+(-1.25,1)$) circle (3pt);
    \filldraw[black] ($(B)+(-1.25,1)$) circle (3pt);
    \filldraw[black] ($(D)+(-1.25,1)$) circle (3pt);
\end{tikzpicture}\hspace{5mm},\hspace{5mm}
\begin{tikzpicture}[baseline=-0.6ex, scale = 0.75]
    \node[tensorSqua, fill=pastelYellow!90] (A){}; 
    \draw[index] (A) -- ++(-1,0);
    \draw[index] (A) -- ++(1,0);
    \draw[index] (A) -- ++(0,1) node[above]{$\mu$};
\end{tikzpicture}
= \left[\begin{array}{cc}
    1 & 0 \\
    1-\delta_{\mu,0} & 0
\end{array}\right] \,,
\end{equation*}
leading to the following tensor-network contraction to compute the mass average
\begin{equation*}
m_{\hat{\mathcal{O}}} = 
\begin{tikzpicture}[baseline=-0.6ex, node distance=1cm, scale=0.75]
    \node[tensorSqua, fill=pastelYellow!90] (A) {};
    \node[tensorSqua, fill=pastelYellow!90, right=of A] (B) {};
    \node[draw=none, right=of B] (C) {};
    \node[tensorSqua, fill=pastelYellow!90, right=of C] (D) {};

    \draw[index] (A) -- (B);
    \draw[index] (B) -- ++(1,0);
    \draw[index, dashed] ($(B)+(1.25,0)$) -- ($(D)+(-1.25,0)$);
    \draw[index] (D) -- ++(-1,0);

    \draw[index] (A) -- ++(0,1);
    \draw[index] (B) -- ++(0,1);
    \draw[index] (D) -- ++(0,1);

    \draw[index] ($(A)+(0,1)$) -- ++($(-1.25,0)$);
    \draw[index] ($(B)+(0,1)$) -- ++($(-1.25,0)$);
    \draw[index] ($(D)+(0,1)$) -- ++($(-1.25,0)$);

    \draw[index] ($(A)+(-1.25,1.5)$) -- ++(0,-3);
    \node[tensorCir, fill=tensorTeal!80, above=of {$(A)+(-1.25,0.15)$}](N1u) {};
    \node[tensorCir, fill=tensorTeal!80, above=of {$(A)+(-1.25,-3.5)$}](N1d) {};
    \draw[index] ($(B)+(-1.25,1.5)$) -- ++(0,-3);
    \node[tensorCir, fill=tensorTeal!80, above=of {$(B)+(-1.25,0.15)$}](N2u) {};
    \node[tensorCir, fill=tensorTeal!80, above=of {$(B)+(-1.25,-3.5)$}](N2d) {};
    \draw[index] ($(D)+(-1.25,1.5)$) -- ++(0,-3);
    \node[tensorCir, fill=tensorTeal!80, above=of {$(D)+(-1.25,0.15)$}](N3u) {};
    \node[tensorCir, fill=tensorTeal!80, above=of {$(D)+(-1.25,-3.5)$}](N3d) {};

    \draw[index] (N1u) -- (N2u);
    \draw[index] (N2u) -- ++(1,0);
    \draw[index] (N3u) -- ++(-1,0);
    \draw[index, dashed] ($(N2u)+(1.25,0)$) -- ($(N3u)+(-1.25,0)$);

    \draw[index] (N1d) -- (N2d);
    \draw[index] (N2d) -- ++(1,0);
    \draw[index] (N3d) -- ++(-1,0);
    \draw[index, dashed] ($(N2d)+(1.25,0)$) -- ($(N3d)+(-1.25,0)$);

    \filldraw[black] ($(A)+(-1.25,1)$) circle (3pt);
    \filldraw[black] ($(B)+(-1.25,1)$) circle (3pt);
    \filldraw[black] ($(D)+(-1.25,1)$) circle (3pt);
\end{tikzpicture} \,.
\end{equation*}
The full mass probability distribution $P_{\hat{\mathcal{O}}}(m)$ can be extracted from the spectral
decomposition of $\hat{M}$.
A convenient approach is to express the projector onto the eigenspace with eigenvalue $m$
as a discrete Fourier transform:
\begin{equation}
    P_{\hat{\mathcal{O}}}(m)
    = \langle \hat{\mathcal{O}} | \delta_{\hat{M},m} | \hat{\mathcal{O}} \rangle
    = \frac{1}{L+1}
      \sum_{k=0}^{L} e^{-i\alpha k m} \,
      G_{\hat{\mathcal{O}}}(\alpha k),
    \qquad
    \alpha = \frac{2\pi}{L+1},
\end{equation}
where
\begin{equation}
    G_{\hat{\mathcal{O}}}(\lambda)
    = \Big\langle \hat{\mathcal{O}} \Big|
      \prod_{j=1}^{L}
      e^{\,i\lambda(\hat{\mathbb{I}}_j - |0\rangle\langle 0|_j)}
      \Big| \hat{\mathcal{O}} \Big\rangle
\end{equation}
acts as a generating function for the moments of the mass operator. In particular, $G_{\hat{\mathcal{O}}}(\alpha k)$ can be evaluated iteratively in $k$, within the Pauli MPS framework, as follows
\begin{equation*}
G_{\hat{\mathcal{O}}}(\alpha k) = 
\begin{tikzpicture}[baseline=-5ex, node distance=1cm, scale=0.75]
    \node[tensorSqua, fill=pastelPink!60] (A) {};
    \node[tensorSqua, fill=pastelPink!60, right=of A] (B) {};
    \node[draw=none, right=of B] (C) {};
    \node[tensorSqua, fill=pastelPink!60, right=of C] (D) {};

    \node[tensorSqua, fill=pastelPink!60, below=15pt of A] (A1) {};
    \node[tensorSqua, fill=pastelPink!60, right=of A1] (B1) {};
    \node[draw=none, right=of B1] (C1) {};
    \node[tensorSqua, fill=pastelPink!60, right=of C1] (D1) {};

    \draw[index, dashed] ($(B)+(1,0)$) -- ($(D)+(-1,0)$);
    \draw[index, dashed] ($(B1)+(1,0)$) -- ($(D1)+(-1,0)$);

    \draw[index] (A) -- ++(0,1);
    \draw[index] (B) -- ++(0,1);
    \draw[index] (D) -- ++(0,1);

    \draw[index] ($(A)+(0,1)$) -- ++($(-1.25,0)$);
    \draw[index] ($(B)+(0,1)$) -- ++($(-1.25,0)$);
    \draw[index] ($(D)+(0,1)$) -- ++($(-1.25,0)$);

    \draw[index] (A1) -- ++(0,1);
    \draw[index] (B1) -- ++(0,1);
    \draw[index] (D1) -- ++(0,1);
    
    \draw[index] ($(A1)+(0,1)$) -- ++($(-1.25,0)$);
    \draw[index] ($(B1)+(0,1)$) -- ++($(-1.25,0)$);
    \draw[index] ($(D1)+(0,1)$) -- ++($(-1.25,0)$);

    \draw[index] ($(A)+(-1.25,1.5)$) -- ++(0,-3.5);
    \node[tensorCir, fill=tensorTeal!80, above=of {$(A)+(-1.25,0.15)$}](N1u) {};
    \node[tensorCir, fill=tensorTeal!80, above=of {$(A)+(-1.25,-6.5)$}](N1d) {};
    \draw[index] ($(B)+(-1.25,1.5)$) -- ++(0,-3.5);
    \node[tensorCir, fill=tensorTeal!80, above=of {$(B)+(-1.25,0.15)$}](N2u) {};
    \node[tensorCir, fill=tensorTeal!80, above=of {$(B)+(-1.25,-6.5)$}](N2d) {};
    \draw[index] ($(D)+(-1.25,1.5)$) -- ++(0,-3.5);
    \node[tensorCir, fill=tensorTeal!80, above=of {$(D)+(-1.25,0.15)$}](N3u) {};
    \node[tensorCir, fill=tensorTeal!80, above=of {$(D)+(-1.25,-6.5)$}](N3d) {};

    \draw[index] (N1u) -- (N2u);
    \draw[index] (N2u) -- ++(1,0);
    \draw[index] (N3u) -- ++(-1,0);
    \draw[index, dashed] ($(N2u)+(1.25,0)$) -- ($(N3u)+(-1.25,0)$);

    \draw[index] (N1d) -- (N2d);
    \draw[index] (N2d) -- ++(1,0);
    \draw[index] (N3d) -- ++(-1,0);
    \draw[index, dashed] ($(N2d)+(1.25,0)$) -- ($(N3d)+(-1.25,0)$);

    \filldraw[black] ($(A)+(-1.25,1)$) circle (3pt);
    \filldraw[black] ($(B)+(-1.25,1)$) circle (3pt);
    \filldraw[black] ($(D)+(-1.25,1)$) circle (3pt);

    \filldraw[black] ($(A1)+(-1.25,1)$) circle (3pt);
    \filldraw[black] ($(B1)+(-1.25,1)$) circle (3pt);
    \filldraw[black] ($(D1)+(-1.25,1)$) circle (3pt);

    \draw[index] (N1d) -- ++(0,1);
    \draw[index] (N2d) -- ++(0,1);
    \draw[index] (N3d) -- ++(0,1);
    \draw[index, dashed] ($(A)+(-1.25,-2.25)$) -- ++(0,-1);
    \draw[index, dashed] ($(B)+(-1.25,-2.25)$) -- ++(0,-1);
    \draw[index, dashed] ($(D)+(-1.25,-2.25)$) -- ++(0,-1);

    \draw[decorate, decoration={brace, amplitude=5pt}, very thick]($(N3u)+(2.25,-0.75)$) -- ($(N3d)+(2.25,0.75)$)
    node[midway, right=4pt] {$\boldsymbol{k}$};
\end{tikzpicture}\hspace{5mm},\hspace{5mm}
\begin{tikzpicture}[baseline=-0.5ex, scale=0.75]
    \node[tensorSqua, fill=pastelPink!60] (A){};
    \draw[index] (A) -- ++(0,1);
\end{tikzpicture} = e^{i\alpha(1-\delta_{\mu,0})} \,.
\end{equation*}
The iterative evaluation of $G_{\hat{\mathcal{O}}}(\lambda)$ in the MPS representation scales as $\mathcal{O}(L^2 \chi^3)$.

Physically, the operator mass measures the number of sites
that actively participate in the operator at a given time.
While the operator length $h_{\hat{\mathcal{O}}}$ discussed in Sec.~\ref{olen:sec} captures the farthest site reached by $\hat{\mathcal{O}}$,
the mass $m_{\hat{\mathcal{O}}}$ quantifies its internal structure:
it distinguishes, for instance, a sparse string of few nontrivial spins from a dense operator filling the entire region up to $h_{\hat{\mathcal{O}}}$. 
\textcolor{red}{For instance, the operator in Eq.~(\ref{eq::examplesOp}) is represented as a superposition of basis states with support on different sites, which can be efficiently encoded as a MPS. The MPO representation of $\hat{\mathcal{O}}$ thus translates into an MPS describing the weight of different Pauli strings, allowing us to compute observables such as $m$ and $h$ by evaluating appropriate expectation values in this extended space.}


\subsection{Operator entanglement entropy}\label{oen:sec}

A complementary probe of the operator growth is the \emph{operator entanglement 
entropy}, introduced in~\cite{bandyopadhyay2005entanglingpowerquantumchaotic,PhysRevA.76.032316,PhysRevB.79.184416}, and defined as the bipartite entanglement of the Choi--Jamiołkowski state Eq.~\eqref{statop:eqn} associated with the operator~\cite{choi1975completely,jamiolkowski1972linear}. While the operator mass and operator length quantify the spatial support of $\hat O(t)$ in the Pauli basis, the operator entanglement probes the \emph{complexity} of the operator within that support, i.e., the pattern of entanglement it develops across spatial bipartitions.
\paragraph{Bipartition of the operator state ---}

We consider a spatial bipartition of the system into 
$A = \{1,\ldots,\ell\}$ and $B=\{\ell+1,\ldots,L\}$, which induces a bipartition 
of the vectorized operator $|\hat O\rangle$.  The reduced density 
matrix associated to subsystem $A$ is
\begin{equation}
\rho_A = \Tr_B\! \big[ |\hat O\rangle \langle \hat O| \big],
\end{equation}
and the corresponding operator entanglement entropy is
\begin{equation}
	e_{\ell} \equiv S_{\mathrm{op}}(\ell) = -\Tr \big[ \rho_A \log \rho_A \big].
\end{equation}
In numerical simulations based on tensor networks, it is often advantageous to use the second-order Rényi entropy,
\begin{equation} \label{lindo:eqn}
  S^{(2)}_{\mathrm{op}}(\ell) = -\log\!\left[\Tr(\rho_A^2)\right],
\end{equation}
which can be obtained directly from the Schmidt coefficients without explicitly 
constructing $\rho_A$.
\paragraph{Evaluation from the operator MPS ---}

When $|\hat O\rangle$ is represented as a MPS in the
canonical form, its Schmidt decomposition across the $\ell$ bond is
\begin{equation}
|\hat O\rangle = \sum_{a=1}^{\chi_\ell} 
\lambda^{(\ell)}_a\, |a\rangle_{A} \otimes |a\rangle_{B},
\end{equation}
where $\{\lambda^{(\ell)}_a\}$ are the Schmidt singular values associated with the bipartition. In this representation, the operator entanglement is
\begin{align} \label{lindo1:eqn}
  e_{\ell} = S_{\mathrm{op}}(\ell)
&= - \sum_a (\lambda^{(\ell)}_a)^{2}\,
   \log \! \Big[(\lambda^{(\ell)}_a)^{2}\Big], \\
S^{(2)}_{\mathrm{op}}(\ell)
&= -\log\! \bigg[ \sum_a (\lambda^{(\ell)}_a)^4 \bigg].
\end{align}
Thus the operator entanglement entropy directly reflects the bond spectrum and 
provides a quantitative measure of the bond dimension required to represent 
$\hat O(t)$ faithfully in an MPS-based time evolution such as TEBD or tDMRG.
\paragraph{Physical considerations ---}

The operator entanglement characterizes aspects of the operator growth that are 
invisible to quantities based solely on spatial support.  
The operator length $h_{\hat O}(t)$ measures the farthest site reached by 
nonzero operator weight, and the operator mass $m_{\hat O}(t)$ quantifies how 
many sites contribute significantly.  
By contrast, the operator entanglement entropy captures the structure of 
entanglement generated within that region, providing a measure of the 
many-body complexity of $\hat O(t)$ as a state in the doubled Hilbert space.

In chaotic systems, $S_{\mathrm{op}}(t)$ of an initially local Hermitian operator typically grows linearly at early times, 
with a rate set by the operator entanglement velocity~\cite{jonay2018coarsegraineddynamicsoperatorstate,Bertini_2020}.  
The entropy eventually saturates to a volume-law value determined by the 
system size.  
In integrable dynamics, operator entanglement grows more slowly -- often 
logarithmically~\cite{Muth_2011,PhysRevA.76.032316,PhysRevB.79.184416,PhysRevLett.122.250603} -- reflecting the existence of an extensive number of
local operators conserved by the dynamics~\cite{Essler_2016}. This difference is related to the fact that generic evolving
quantum operators are represented with MPO with a rank increasing exponentially in time, while the rank
increases linearly in time for integrable systems~\cite{PhysRevE.75.015202}.

Operator entanglement has also been considered for the unitary time evolution operator itself~\cite{Dubail_2017}, and it has
been predicted that it linearly increases in time unless the system is in a localized phase. This ballistic increase has been confirmed
in driven nonintegrable systems, while in the disordered XXZ chain in the ergodic regime the increase is power-law with exponent $1/2$ (diffusive).
Most remarkably the increase is logarithmic in the MBL regime~\cite{PhysRevB.95.094206}. As far as we know, in the disordered XXZ chain there are no studies at present
of the operator entanglement evolution of an initially localized Hermitian operator, and in this work we are going to fill this gap.
%


In the next section, we study the evolution of the quantities introduced in this section in the unitary dynamics of a disordered spin chain.
%
\section{Operator spreading in an interacting disordered spin chain}\label{sec:results}
\subsection{The model}
We consider the unitary dynamics of the disordered spin-1/2 Hamiltonian
\begin{equation}
\label{hmbl:eqn}
  \hat{H} = \sum_{j=1}^{L-1}\left[\frac{J}2\left(\hat{\sigma}_j^+\hat{\sigma}_{j+1}^-+\mbox{H.c.}\right) + \frac{\Delta}{4}\hat{\sigma}_j^z\hat{\sigma}_{j+1}^z\right] + \sum_{j=1}^L\frac{h_j}{2}\hat{\sigma}_j^z \,,
\end{equation}
where $\hat \sigma^\pm_j = \tfrac12(\hat \sigma^x_j \pm i \hat \sigma^y_j)$ and we consider open boundary conditions. Here $J$ denotes the amplitude of the nearest-neighbor hopping, $\Delta$ is the density-density interaction strength, while the onsite fields $h_j$ are randomly and uniformly distributed in the interval $[-W,W]$. In the following, we set the energy scale by fixing $J=1$.

We study the operator dynamics in the Heisenberg representation, taking as time-zero operator the following one
\begin{equation}\label{opo:eqn}
  \hat{\mathcal{O}} = \hat{\sigma}_1^z \otimes \hat {\mathbb{I}}_2 \otimes \cdots \otimes \hat{\mathbb{I}}_L\,,
\end{equation}
that acts nontrivially only on the first site. We evaluate the operator length and the operator mass, as defined in Sec.~\ref{sec:definitions}, for the Heisenberg-evolved operator at a generic time $t$
\begin{equation}
  \hat{\mathcal{O}}(t) = \nep^{i\hat{H} t} \hat{\mathcal{O}} \nep^{-i\hat{H} t}\,,
\end{equation}
thereby studying its spreading on the basis of the Pauli operators. We emphasize the time dependence calling $h(t)$ the length of this operator and $m(t)$ its operator mass. We then average over $N_{\rm r}$ disorder realizations, marking disorder averages with an overline $\overline{(\cdots)}$.
\begin{figure}[t!]
 \centering
 \includegraphics[width=0.8\textwidth]{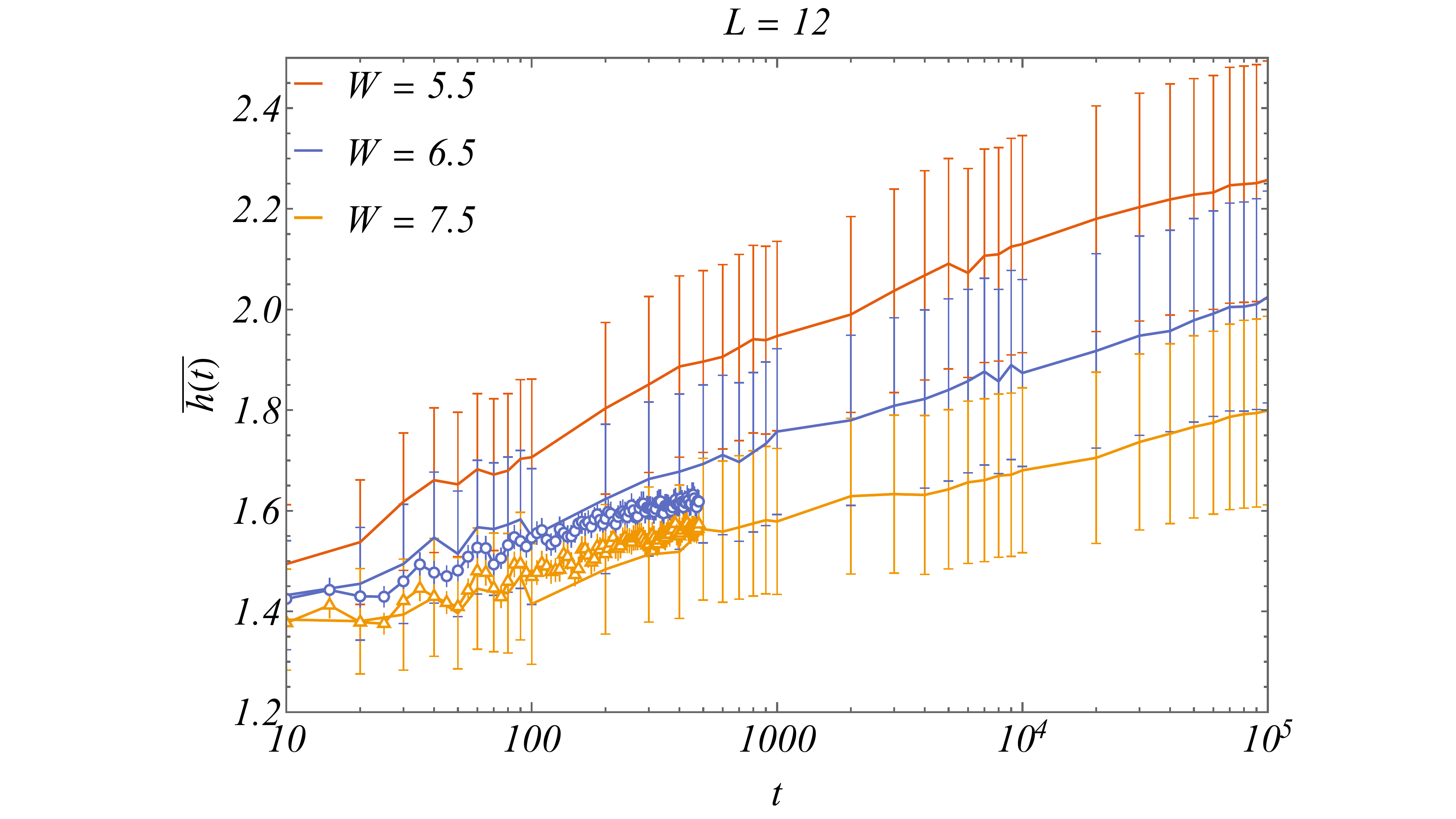}\\
 \caption{\textcolor{red}{The average operator length $\overline{h}(t)$ as a function of time $t$ for the model of Eq.~\eqref{hmbl:eqn}, for a system with $L = 12$ sites. Solid lines correspond to ED results, reaching times up to $t \sim 10^{5}$, while symbols denote TN data, available up to $t \sim 5 \times 10^{2}$. The TN results are overlaid on the ED data in the time window where both methods overlap, showing good agreement within the errorbars, and thereby benchmarking the tensor-network approach. Here we fix $\Delta = 1$, while different curves correspond to different disorder strengths $W$ (see legend). Averages are performed over $N_r = 48$ disorder realizations for ED and $N_r = 200$ for TN. Notice the logarithmic scale on the horizontal axis and the logarithmic growth of $\overline{h}(t)$ over several decades.}}
 \label{fig:h_ED}
\end{figure}
\begin{figure}
\centering
\begin{tabular}{l}
(a)\\ 
\includegraphics[width=0.8\textwidth]{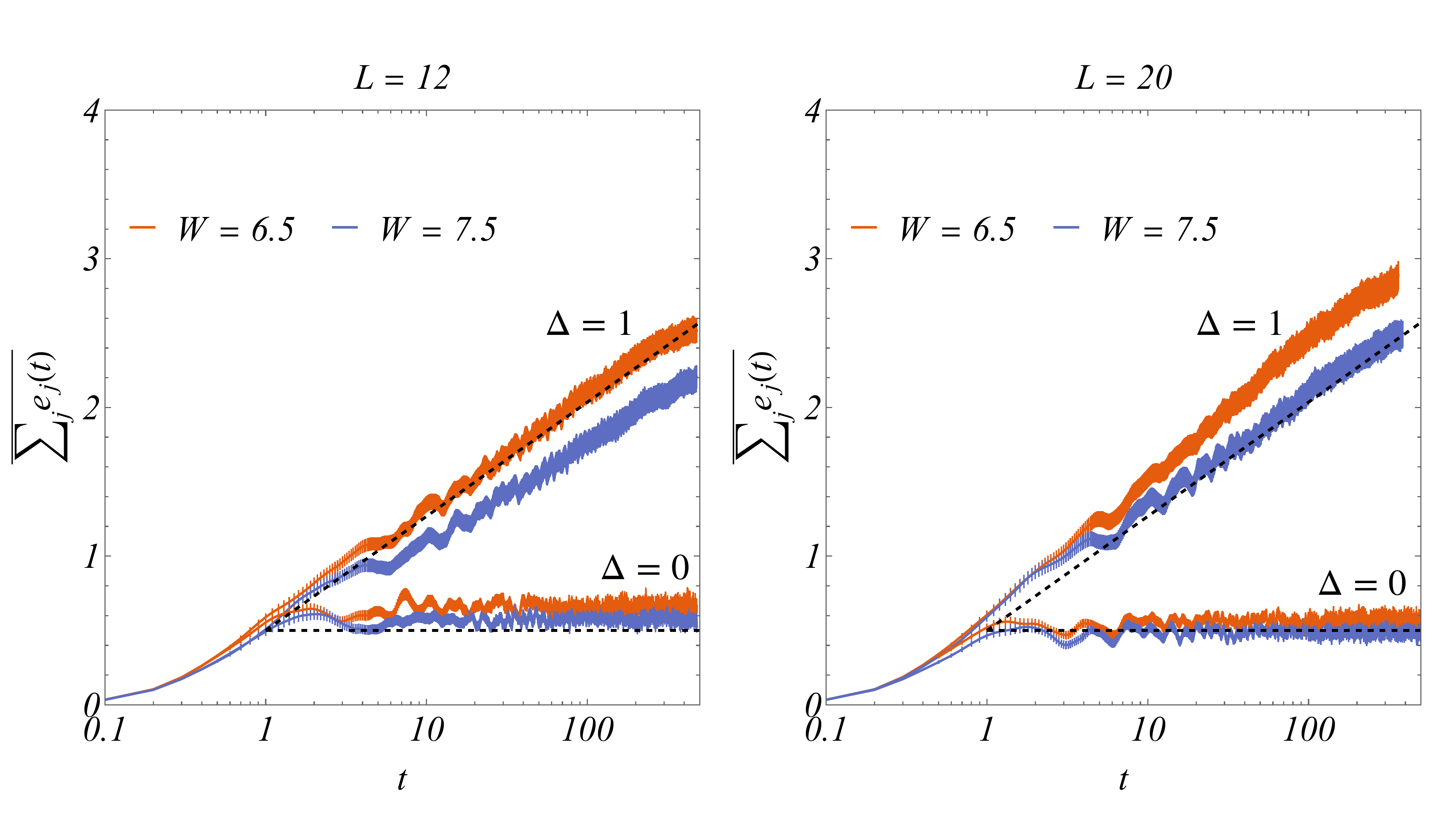}\\
(b)\\
\includegraphics[width=0.8\textwidth]{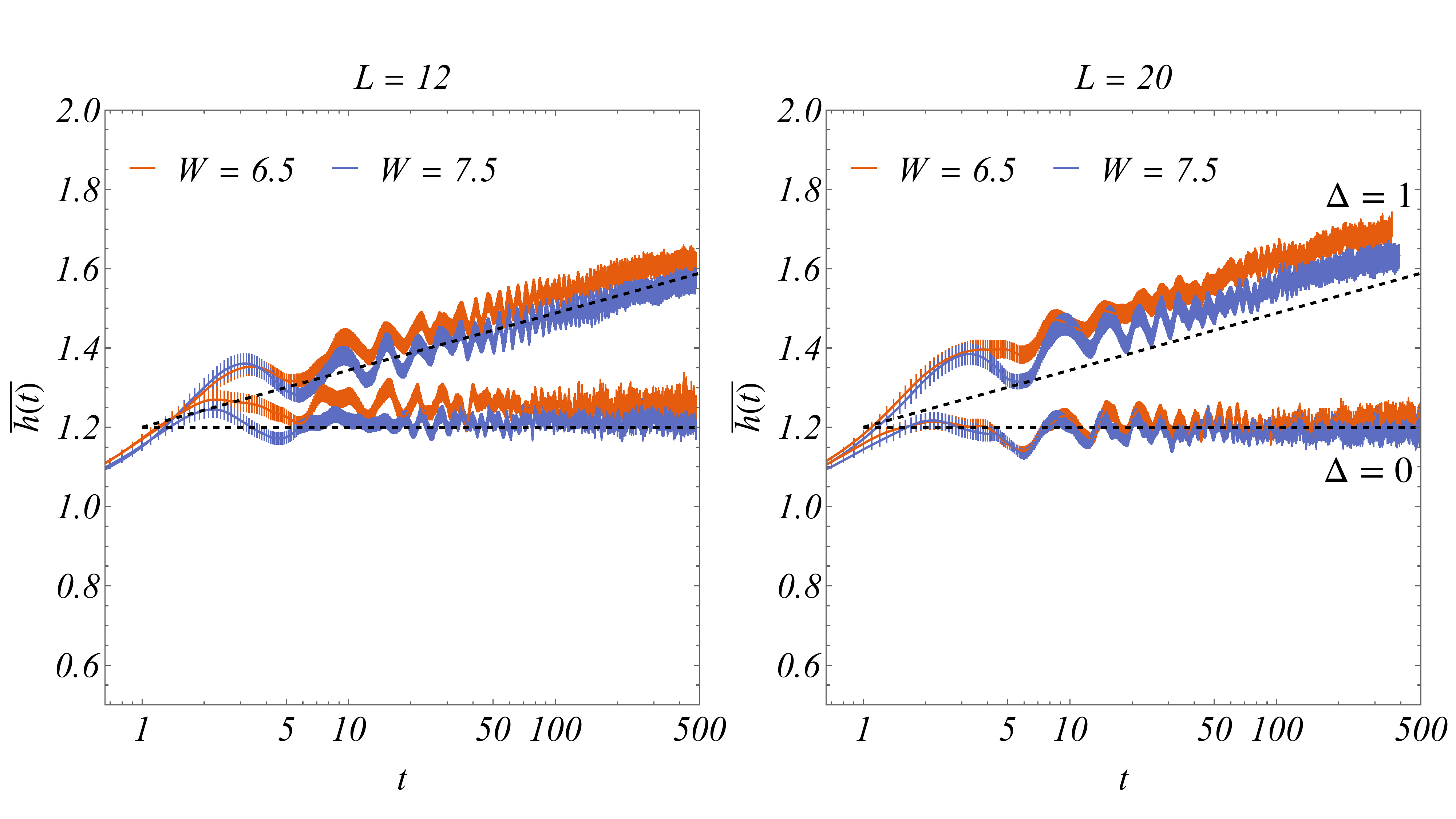}\\
(c)\\
\includegraphics[width=0.8\textwidth]{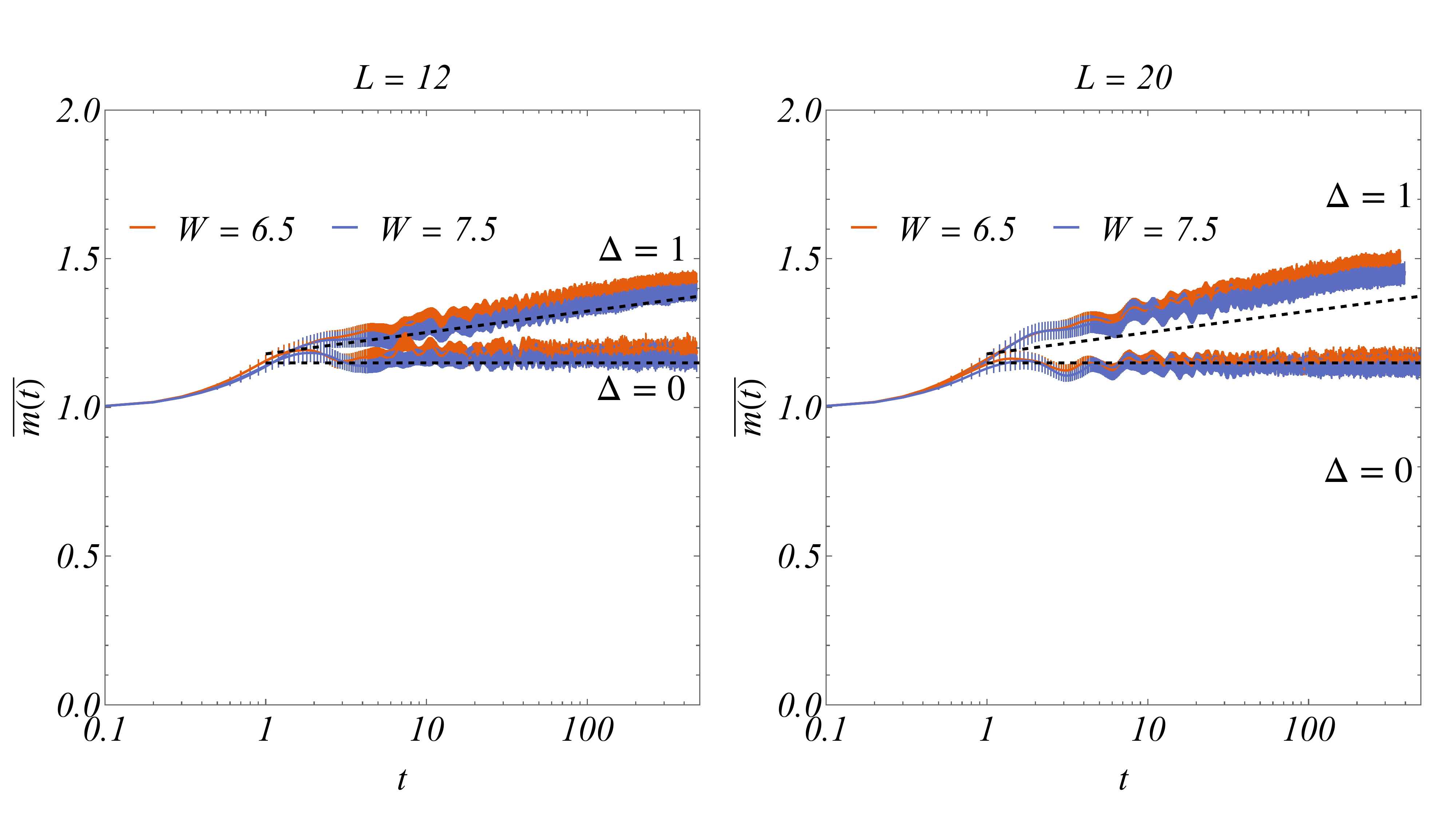}
\end{tabular}
\caption{MPS dynamics of the integrated operator entanglement (a), the operator length $\overline{h}(t)$ (b), and the operator mass $\overline{m}(t)$ (c), for $L=12$ and $L=20$, up to $t=500$. Notice the logarithmic increase occurring only for the interacting case $\Delta=1$ (dashed lines are guide for the eyes). Averages have been taken over $N_{\rm r} = 200$ disorder realizations.}
\label{fig:mps_ent_h_m}
\end{figure}
%
%
\begin{figure}
    \centering
    \begin{tabular}{l}
(a)\\ 
\includegraphics[width=0.8\textwidth]{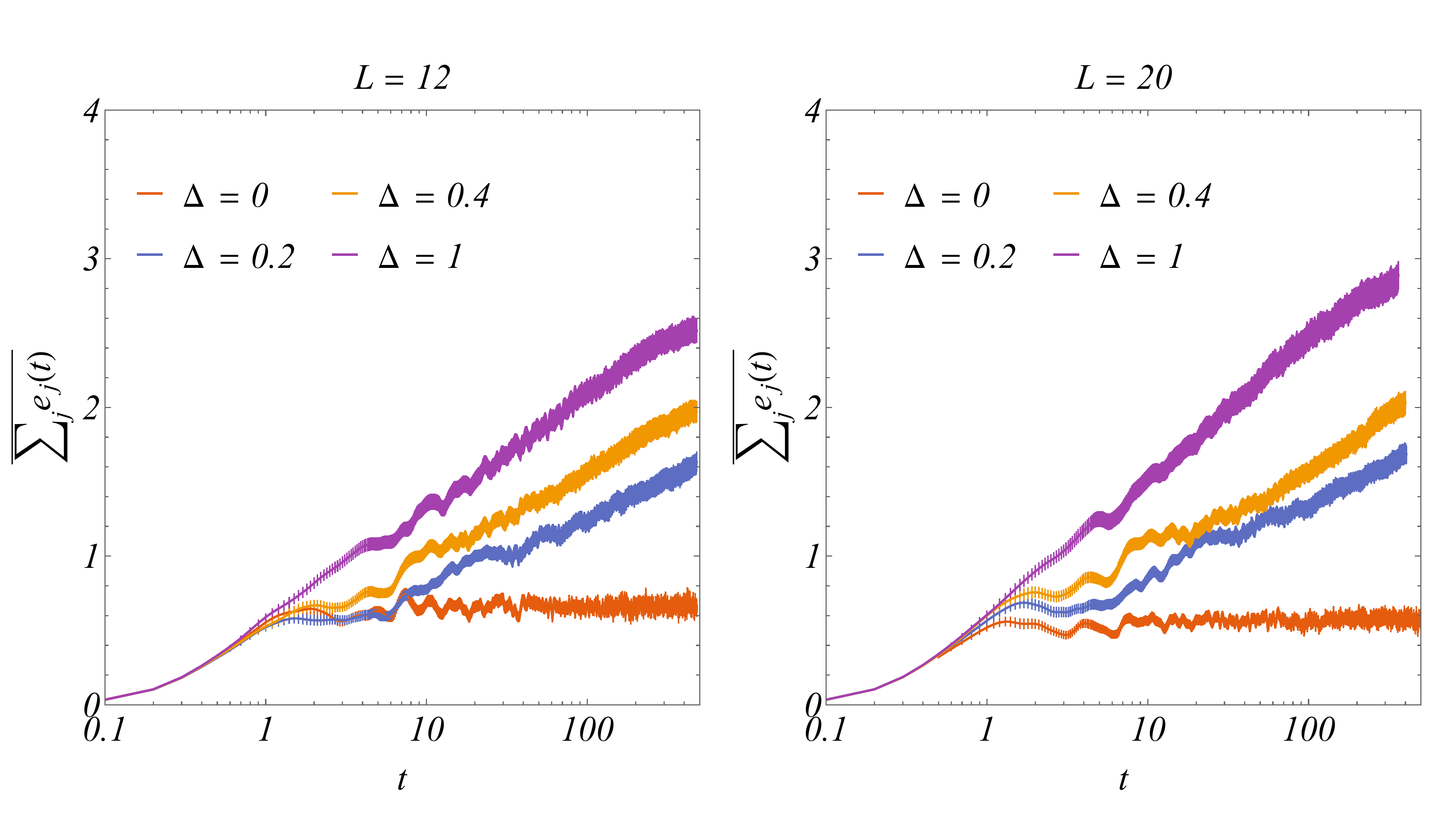}\\
(b)\\
\includegraphics[width=0.8\textwidth]{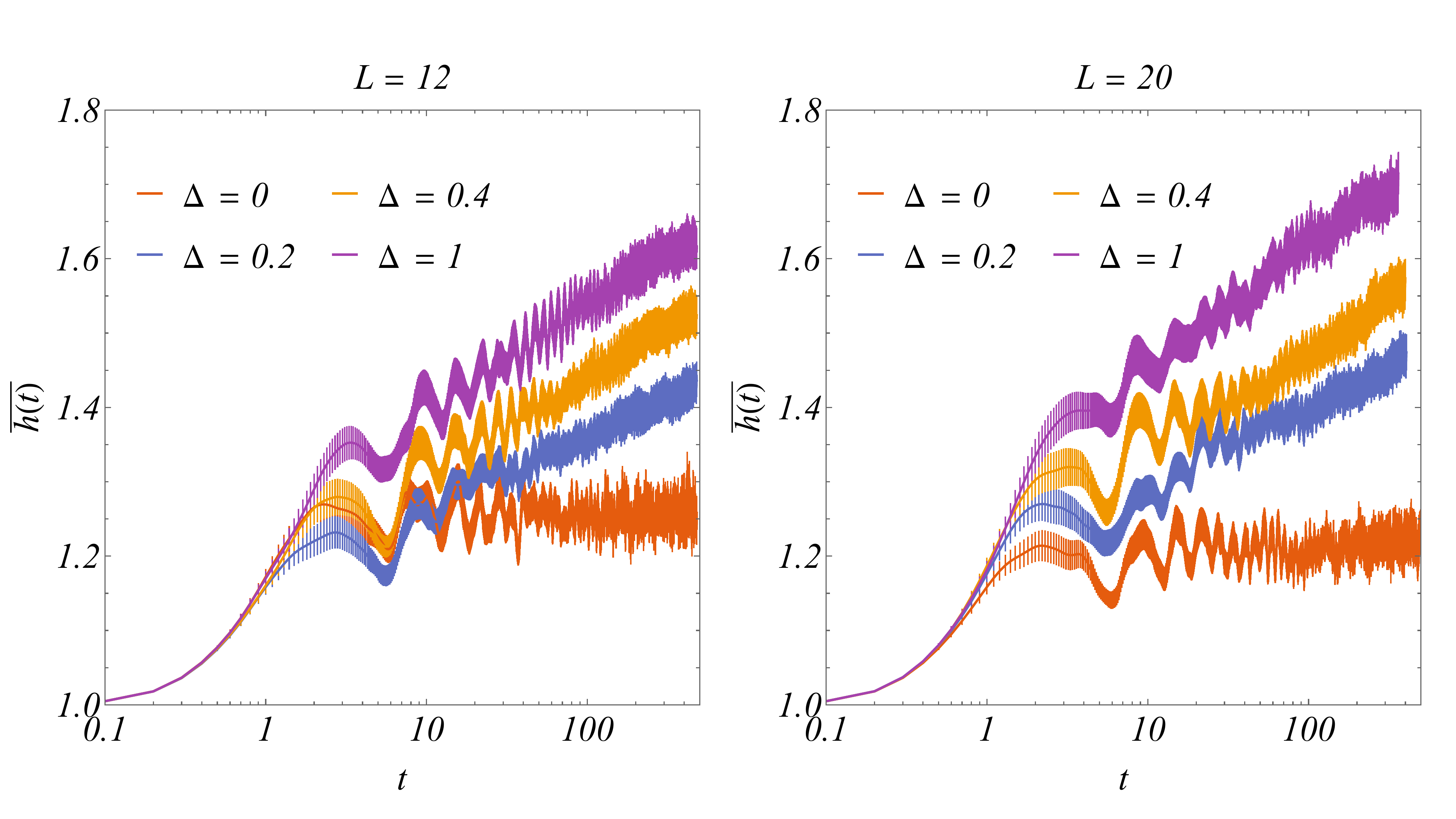}\\
(c)\\
\includegraphics[width=0.8\textwidth]{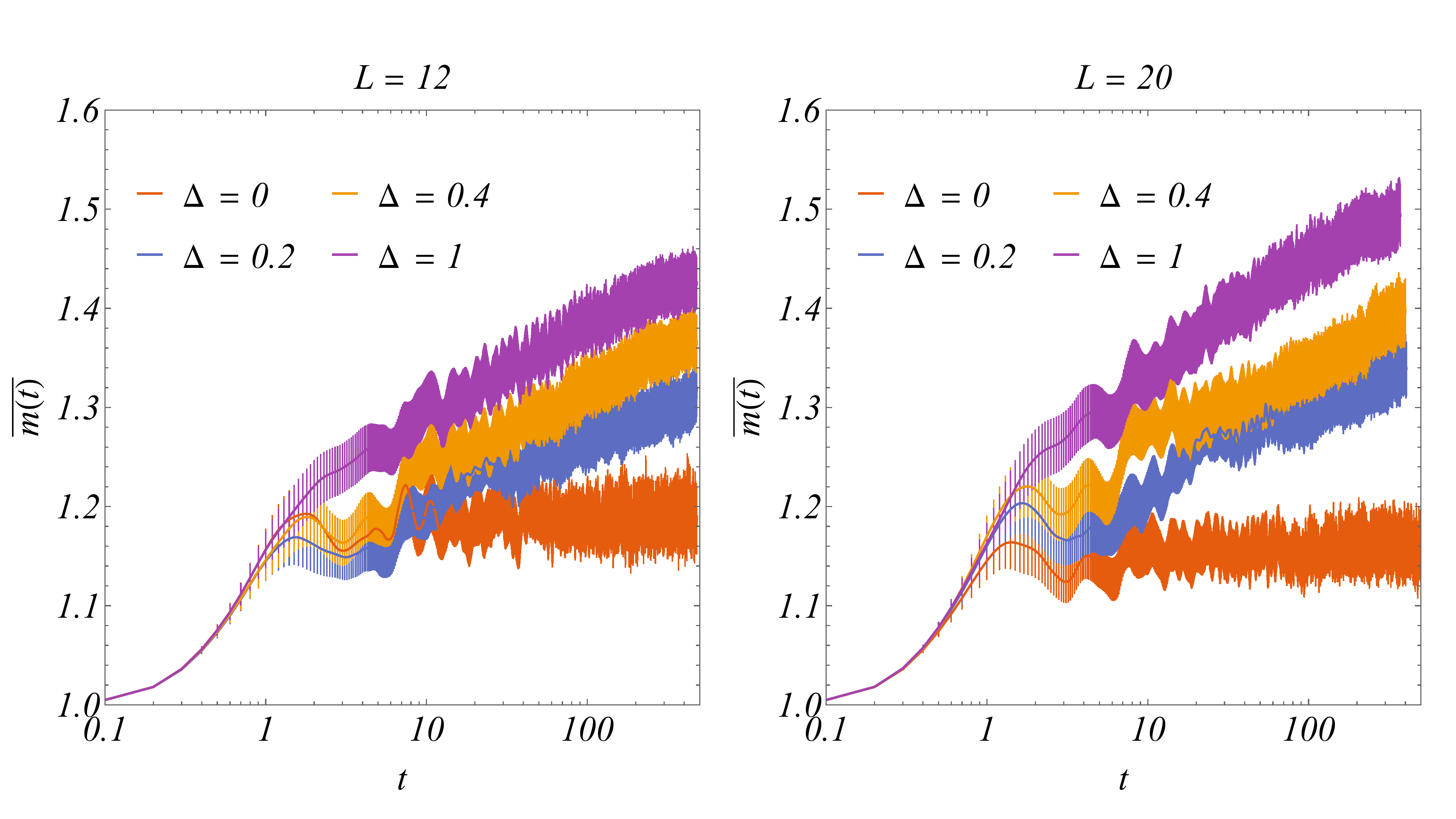}
\end{tabular}
    \caption{Same as in Fig.~\ref{fig:mps_ent_h_m}, but for fixed disorder strength $W = 6.5$ and varying interaction strength $\Delta$. In the non-interacting limit ($\Delta = 0$), all quantities rapidly saturate, signaling the absence of operator spreading. Upon turning on interactions, even weakly, a long-time logarithmic growth emerges abruptly and in a non-perturbative manner, characteristic of the interacting MBL regime.}
    \label{fig:sz_vs_delta}
\end{figure}
%
\subsection{Numerical results}\label{numeris:sec}
To characterize operator spreading in the disordered XXZ chain of Eq.~\eqref{hmbl:eqn}, we resort to exact diagonalization (ED) methods, for $L=12$, and tensor–network simulations based on MPS time evolution, for system sizes up to $L=32$. We focus on three complementary probes of operator growth: (i) the operator length $h(t)$, (ii) the operator mass $m(t)$, and (iii) the operator entanglement entropy $S_{\mathrm{op}}(t)$. As explained in Sec.~\ref{sec:definitions} and~\ref{sec:MPS}, these quantities capture different aspects of operator delocalization: $h(t)$ measures the spatial extent of the operator along  the chain, $m(t)$ quantifies the number of sites on which the operator acts non-trivially, and $S_{\mathrm{op}}(t)$ probes the internal complexity of the operator in the doubled Hilbert space. All results are averaged over disorder realizations, as indicated by the overline notation, and are shown for both interacting ($\Delta\neq 0$) and non–interacting ($\Delta=0$) cases.

\textcolor{red}{To benchmark the tensor-network (TN) approach, we compare its results
with ED for systems of size $L = 12$, where both
methods are accessible. The comparison is shown in
Fig.~\ref{fig:h_ED} for disorder strengths
$W \in \{5.5, \,6.5, \,7.5\}$, at interaction $\Delta = 1$.
The TN results are overlaid on top of the ED data in the time window
where both methods overlap ($t \lesssim 5 \times 10^{2}$), showing good agreement within the errorbars and thereby
validating the accuracy of our TN simulations. At longer
times, only ED results are available, extending the dynamics over
several decades.
The dynamics displays a clear logarithmic growth,
$\overline{h}(t) \sim a \ln t + b$, consistent with the phenomenology
of MBL systems, where operator spreading is governed
by exponentially weak interactions between emergent $\ell$-bits
~\cite{PhysRevLett.110.260601,PhysRevB.90.174202,PhysRevLett.111.127201,Nandkishore_2015}.}

As we show in detail in~\ref{appendix:l-bit}, the $\ell$-bit model predicts the logarithmic increase in time of $\overline{h}(t)$ that we are here numerically observing. Due to the interactions exponentially decaying with distance between the $\ell$-bits, an operator initially localized on site $1$ can only develop support at distance $r$ through an effective coupling $J_{1r}\sim e^{-r/\xi}$, which induces dephasing on timescales $t_r\sim e^{r/\xi}$. Inverting this relation gives $r(t)\sim \xi\ln t$, directly predicting a logarithmic operator length. Our ED data clearly support this picture.
In contrast with this scenario, for $\Delta=0$ the logarithmic growth disappears. In such a case, the system maps through a Jordan-Wigner transformation~\cite{Mbeng_2024} into an Anderson-localized free-fermionic model, where all single-particle eigenstates are localized for $W>0$. Consequently, $\overline{h}(t)$ rapidly saturates to a constant set by the localization length and remains essentially flat for all accessible times, as we numerically find using the MPS formulation [see Fig.~\ref{fig:sz_vs_delta}(b)]

Figure~\ref{fig:mps_ent_h_m} displays the evolution of the integrated operator entanglement (a)~\footnote{Notice that we evaluate the operator entanglement entropy computing the second-order Rényi entropies Eq.~\eqref{lindo:eqn} for all possible values of $\ell$ (that's to say for all possible bipartitions) and summing them, as indicated in the label of the vertical axis in Fig.~\ref{fig:mps_ent_h_m}(a)}, the operator length (b), and the operator mass (c), for $L=12$ (left) and $L=20$ (right), up to $t=500$.
For $\Delta=1$, all the three observables exhibit a clear logarithmic growth in the presence of strong disorder. The logarithmic trend persists across the different system sizes, highlighting the robustness of the MBL-induced slow operator dynamics. The fact that the operator entanglement entropy follows the same scaling as $\overline{h}(t)$ and $\overline{m}(t)$ underscores that the internal complexity of the operator grows at the same slow logarithmic rate as its spatial support -- precisely as suggested by the $\ell$-bit description, where entanglement arises solely from weak dephasing interactions across the chain. This logarithmic increase is reminiscent of the one observed in integrable systems~\cite{Muth_2011,PhysRevA.76.032316,PhysRevB.79.184416,PhysRevLett.122.250603}, and so provides a further confirmation that for the numerically accessible times MBL systems behave in an integrable way. Furthermore, this logarithmic increase means that the MPO rank increases only linearly in time, allowing to perform long-time simulations, precisely as it happens for the MPS rank when simulating the state dynamics in MBL systems~\cite{PhysRevLett.109.017202}.
In contrast, for $\Delta=0$, the operator entanglement remains nearly constant, while both $\overline{h}(t)$ and $\overline{m}(t)$ saturate rapidly. This provides a sharp distinction between Anderson-like localization (free
fermions) and MBL (interacting case): in the free case, even infinitesimal disorder is sufficient to localize all degrees of freedom, preventing any operator growth; in the interacting case, dephasing between $\ell$-bits generates slow but unbounded operator spreading and entanglement growth.

\begin{figure}
 \centering
 \begin{tabular}{l}
  (a) \\
 \includegraphics[width=100mm]{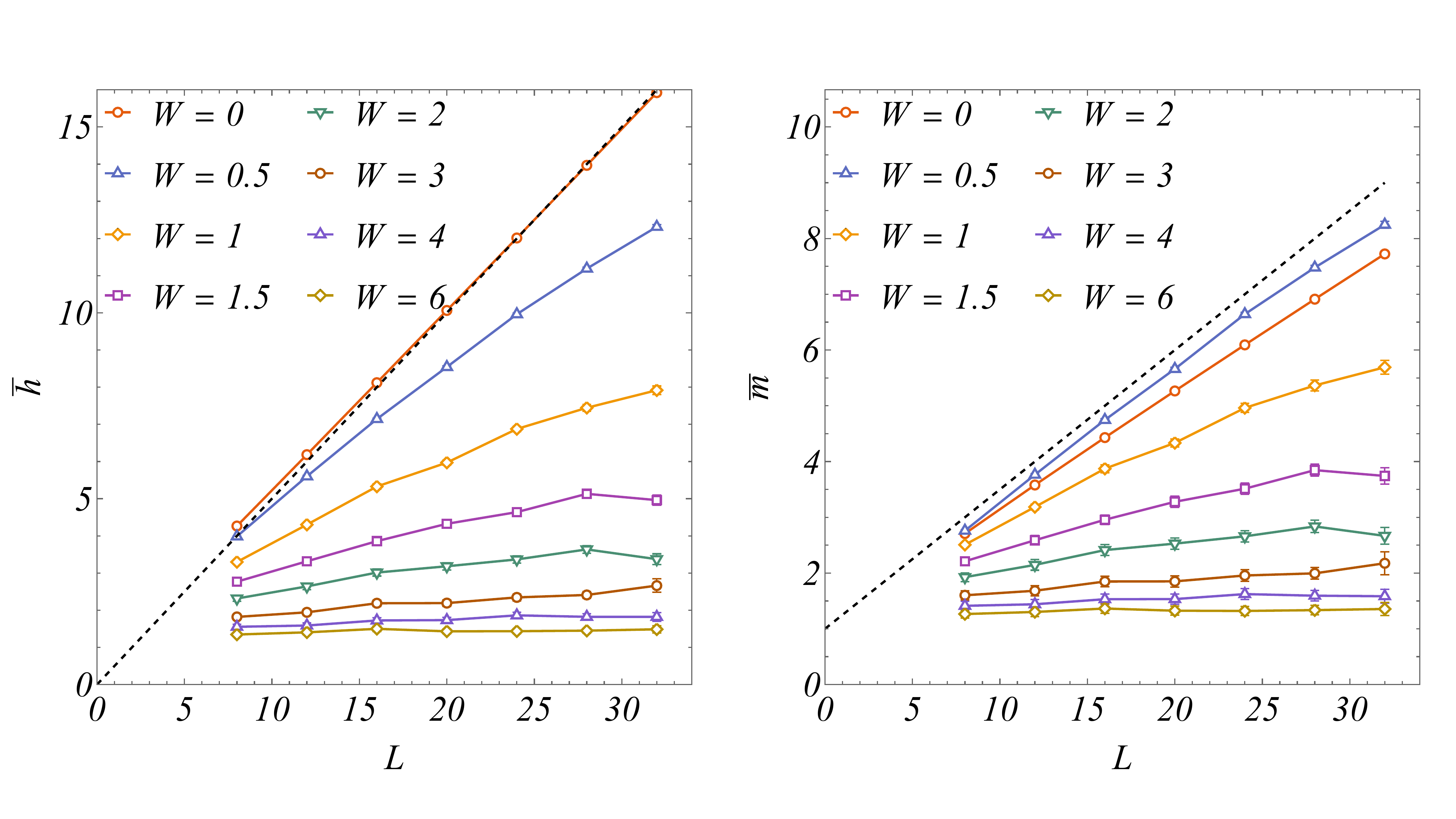}\\
 (b)\\
 \includegraphics[width=100mm]{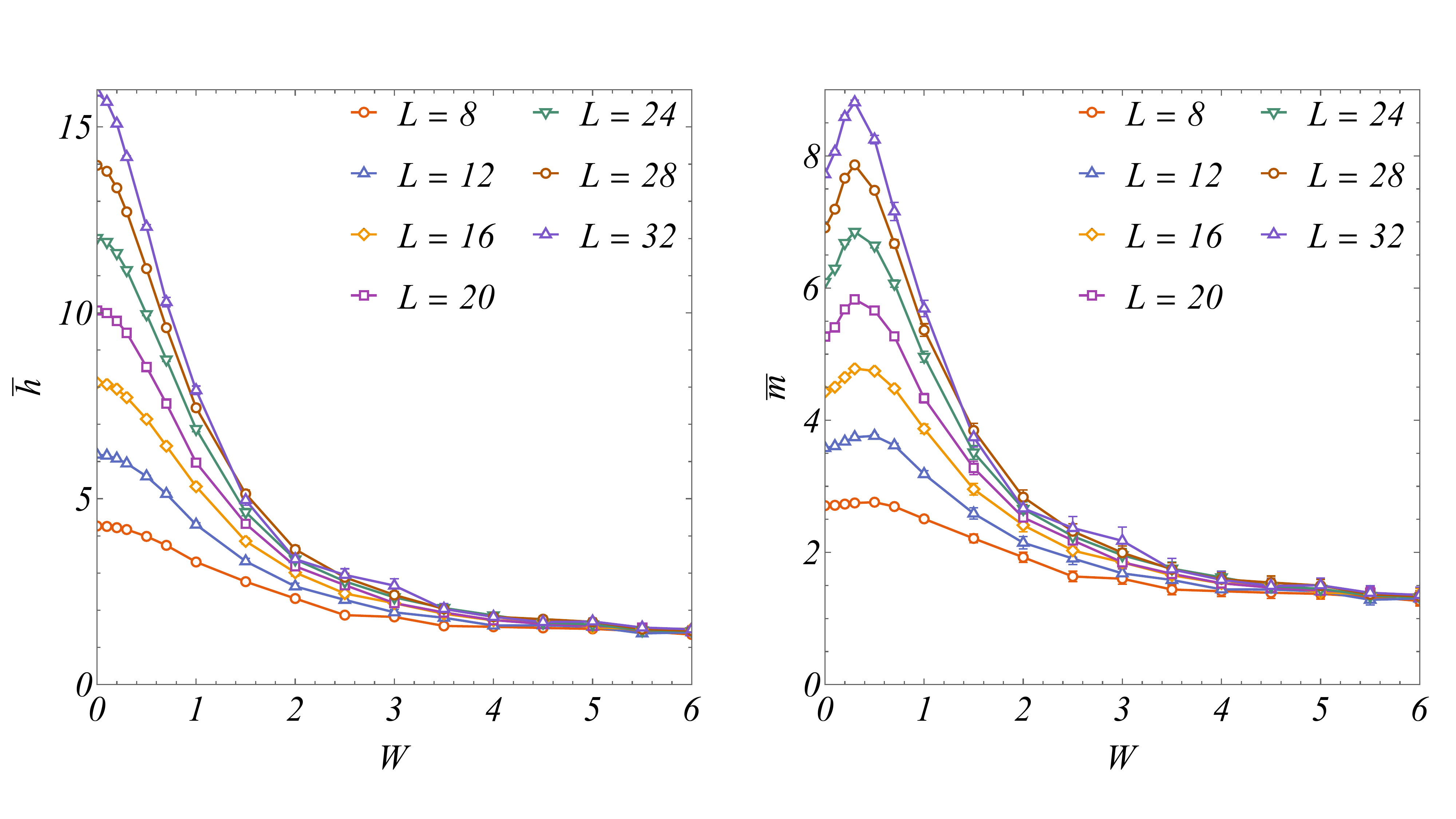}\\
 (c)\\
 \includegraphics[width=100mm]{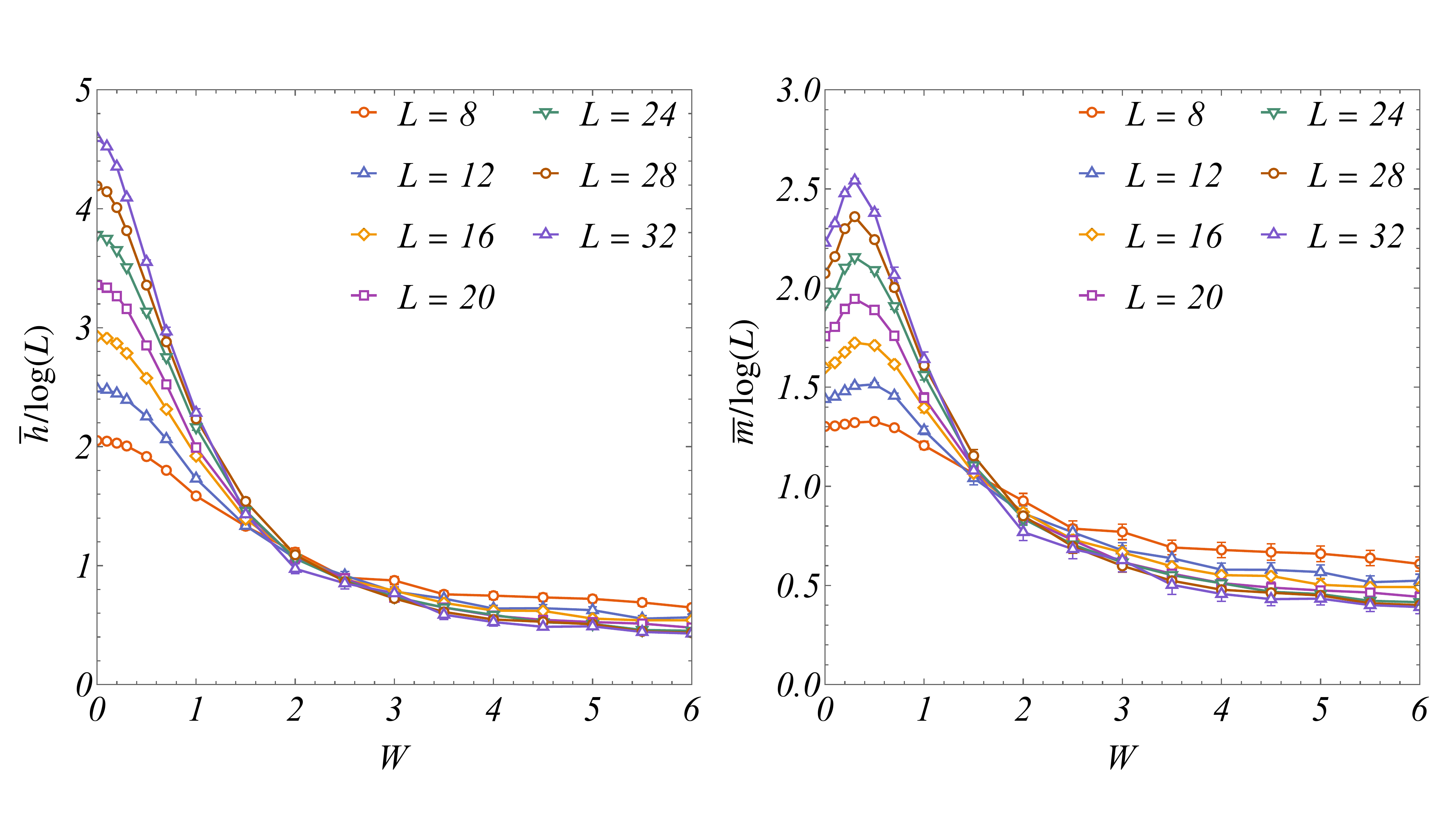}
 \end{tabular}
	\caption{Time-averaged operator length $\overline{h}(L,W)$ and mass $\overline{m}(L,W)$, obtained by integrating $\overline{h}(t,W)$ and $\overline{m}(t,W)$ over the time window $[0,L]$, for $\Delta=1$. Panel (a): dependence on the system size $L$, for different fixed disorder strengths. Panel (b): dependence on the disorder $W$, for fixed system size. Panel (c): rescaled data $\overline{h}(L,W)/\ln L$ and $\overline{m}(L,W)/\ln L$, showing the expected collapse in the many-body localized regime, consistent with the logarithmic light-cone $h(t)\sim \xi(W)\ln t$.}
 \label{logi:fig}
\end{figure}

We now turn to discuss intermediate interaction strengths, $\Delta=0.2$ and $\Delta=0.4$, as shown in Fig.~\ref{fig:sz_vs_delta}. Already for these small but finite interaction values, none of the plotted quantities exhibit a clear saturation at long times. In fact, the operator length, the operator mass, and the operator entanglement entropy all display a logarithmic growth at late times. While the prefactor remains essentially unchanged, the interaction strength primarily modifies sub-leading contributions, resulting in vertical shifts of the curves without altering the underlying logarithmic time dependence. This same behavior is known to occur in the evolution of the state entanglement entropy~\cite{PhysRevLett.109.017202}.
The absence of an extended saturation regime for any $\Delta>0$ indicates that Anderson localization is singular at $\Delta=0$: even arbitrarily weak interactions are sufficient to induce slow dephasing between localized degrees of freedom, leading to unbounded operator spreading and increasing operator complexity at asymptotically long times. This provides further numerical evidence that the interacting term $\Delta$ acts as a relevant perturbation that qualitatively alters the long-time operator dynamics, leading to the logarithmic light-cone behavior characteristic of MBL phases.
%
\paragraph{Time-averaged operator length and mass ---}
To extract the dependence of operator spreading on the system size and the disorder, we define a time average over a window whose length scales with $L$:
\begin{equation}
	\overline{h}(L,W) = \frac{1}{L}\int_0^L \overline{h}(t, W)\,dt ,\qquad
	\overline{m}(L,W) = \frac{1}{L}\int_0^L \overline{m}(t, W)\,dt .
\end{equation}
These quantities probe the ``typical'' operator extent reached within a timescale proportional to the system size. The results are displayed in Fig.~\ref{logi:fig}.

Fig.~\ref{logi:fig}(a) shows $\overline{h}$ and $\overline{m}$ versus the system size. For weak disorder ($W\lesssim 2$), the operator effectively spreads across the whole chain within the window $[0,L]$, resulting in an approximately linear growth with $L$, reminiscent of ergodic or sub-diffusive operator spreading. As $W$ increases, the curves bend downwards and eventually become flat: for strong disorder ($W\gtrsim 6$), the operator remains confined to a small region near its origin for the entire interval $[0,L]$, leading to an $L$-independent value of $\overline{h}$ and $\overline{m}$. The observed flattening matches the physical intuition from the time-trace data: although $h(t)$ logarithmically increases in time at strong disorder, the prefactor $a(W)$ becomes so small that the contribution of the $\ln t$ part to the integral is negligible in the accessible time window, resulting in an essentially constant average.

Figure~\ref{logi:fig}(b) shows the disorder dependence of the averaged quantities for several system sizes. A characteristic non-monotonic trend appears: for weak disorder, operator spreading is enhanced as $W$ increases from zero, because moderate disorder induces ergodicity and suppresses the coherent backscattering of information existing at the integrable point $W=0$; for stronger disorder, localization effects become dominant and the averaged quantities decrease. This non-monotonicity is particularly pronounced in the operator mass, which is more sensitive to the detailed internal structure of the operator than the operator length, and we will better discuss it below. 

Figure~\ref{logi:fig}(c) shows $\overline{h}/\ln L$ and $\overline{m}/\ln L$ as functions of $W$. For strong disorder, the curves collapse onto a nearly $L$-independent profile, consistent with the logarithmic growth $h(t)\approx a(W)\ln t$ derived from the $\ell$-bit model. Indeed, assuming $h(t)\sim a\ln t$, one finds \[ \overline{h}(L,W) \approx a(W)(\ln L -1),\] so that $\overline{h}/\ln L$ directly probes the prefactor $a(W)$. The data collapse in Fig.~\ref{logi:fig}(c) therefore provides a clean numerical signature of the MBL logarithmic light cone.

\textcolor{red}{We note that, at strong disorder, the curves for different system sizes in Fig.~\ref{logi:fig}(b) already appear close to each other, reflecting the weak system-size dependence typical of the deeply localized regime. In this limit, the dynamics is strongly suppressed and the dependence on $L$ becomes mild, leading to an apparent overlap of the data even without any rescaling.
However, this visual proximity should not be interpreted as a genuine scaling collapse. To test the expected scaling form, in Fig.~\ref{logi:fig}(c) we explicitly rescale the data by $\log L$. The resulting collapse provides nontrivial evidence that both $\overline{m}(t)$ and $\bar h(t)$ exhibit a logarithmic dependence on the system size, rather than simply a weak or negligible size dependence. In this sense, the rescaling allows us to distinguish between an apparent overlap and a well-defined scaling behavior. Although the rescaled curves seem to collapse in a range of $W$ approximately corresponding to the one found with standard diagnostics for MBL, the limitations in time and size of our simulations do not allow to exclude that this is just a transient finite-size behavior, as usual for analyses performed with classical-computer limitations~\cite{Sierant_2025_MBL_Class}. Anyway, also a logarithmic light cone as a slow-thermalization transient effect would be an interesting finding; it has been shown in the literature that a logarithmic light cone is not necessarily a consequence of MBL and the presence of extensively many localized integrals of motion~\cite{toniolo2024stabilityslowhamiltoniandynamics}.}

A notable feature of Fig.~\ref{logi:fig} is the non-monotonic dependence of the time-integrated mass on the disorder strength $W$, which exhibits a pronounced peak at intermediate disorder. This phenomenon is probably related to an integrable point at $W=0$ where the system is not space localized but is Bethe-Ansatz integrable~\cite{Franchini_2017}. \textcolor{red}{Such non-monotonic dependence on the disorder strength is already visible in Fig.~\ref{logi:fig}(a), where the time-integrated mass at $W \approx 0.5$
exceeds that at $W=0$ for all system sizes. This anticipates the peak
structure highlighted more clearly in Figs.~4(b) and 4(c).}
\textcolor{red}{We can interpret the physics behind this nonmonotonicity in the following way. At $W=0$ the system is integrable,
thus leading to coherent dynamics in operator space and allowing for partial
reconstruction of simpler operator structures, effectively suppressing
the buildup of operator complexity. Introducing weak disorder breaks
integrability, reduces such coherent backscattering processes, and
enhances the efficiency of operator scrambling, thus leading to an increase
in $\overline{m}$.}

\begin{figure}[!t]
    \centering
    \includegraphics[width=0.9\linewidth]{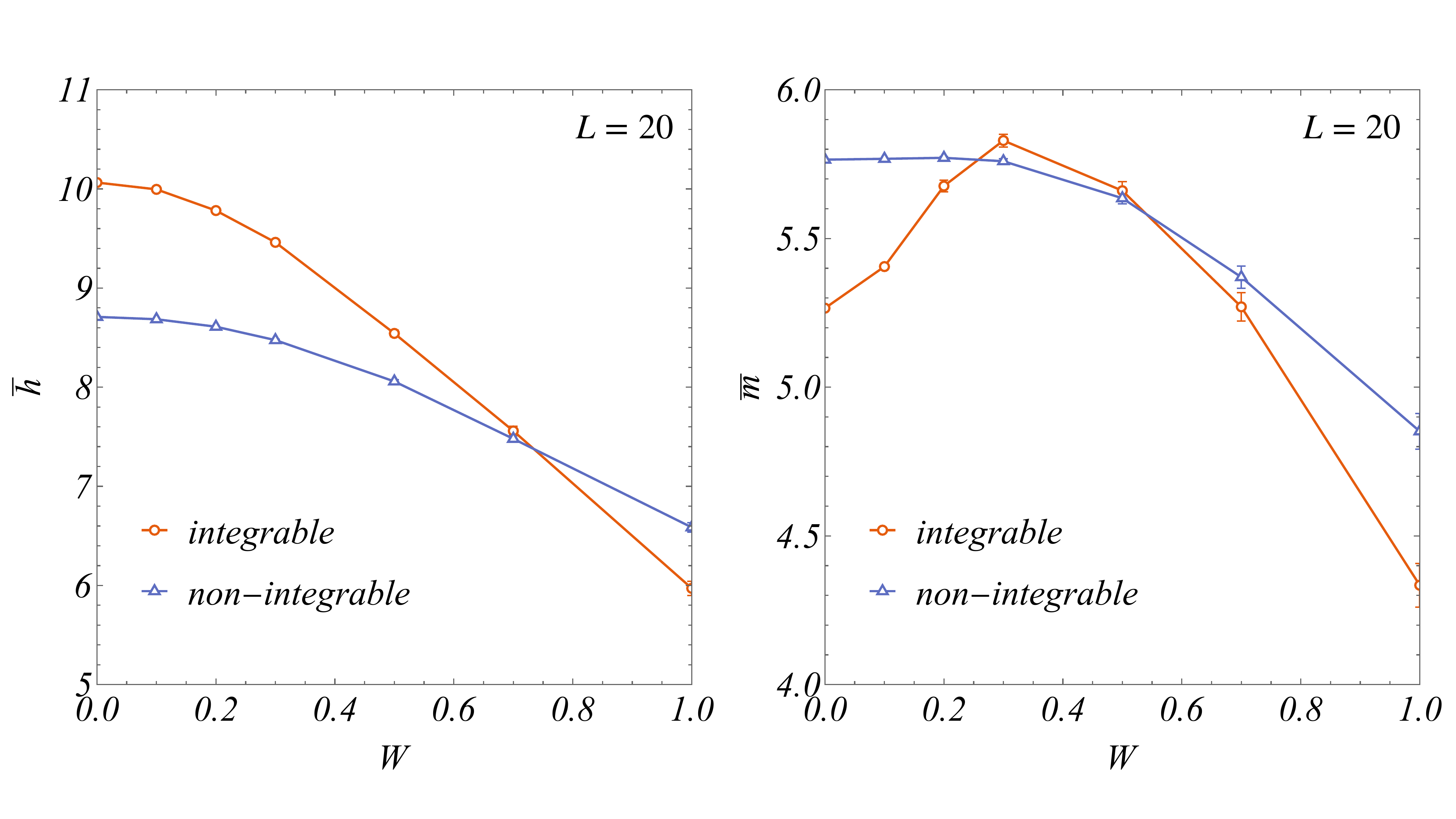}
    \caption{Comparison of the time-integrated operator length $\bar h$ (left) and mass $\overline{m}$ (right) as a function of disorder strength $W$ for XXZ model $(h_x = 0)$ [Eq.~\eqref{hmbl:eqn}] and for the same model with a term that breaks integrability at $W=0$ $(h_x=1)$ [Eq.~\eqref{eq:hbml1}]. The non-monotonic peak observed in the integrable case is reduced when integrability is broken.}
    \label{fig:integrability_breaking}
\end{figure}

\textcolor{red}{To confirm this picture, we have verified that the non-monotonic behavior of $\overline{m}$ is indeed tied to the integrable point at $W=0$. In particular, we add a weak integrability-breaking
perturbation in the form of a transverse external field on the XXZ model
\begin{equation}\label{eq:hbml1}
  \hat{H}' = \hat{H} + h_x\sum_{j=1}^L \hat{\sigma}_j^x\,,
\end{equation}
with $\hat{H}$ given by Eq.~\eqref{hmbl:eqn}. Fixing $\Delta=1/2$, and comparing the cases without and with transverse field $(h_x =1)$, we see that, when integrability at $W=0$ is broken, the non-monotonic
peak in $\overline{m}$ is significantly reduced (see Fig.~\ref{fig:integrability_breaking}, right panel), confirming that the enhancement at small disorder is tied to the proximity to the integrable point.} 

\textcolor{red}{The absence of a non-monotonic behavior in $\overline{h}(L,W)$ versus $W$ can be understood from the fact that it probes only the spatial extent of the operator, i.e., the position of its propagation front (see Fig.~\ref{fig:integrability_breaking}, left panel). This quantity is largely insensitive to the detailed distribution of operator weight and remains controlled by the slow dephasing dynamics characteristic of the MBL regime.
In contrast, the operator mass $\overline{m}(t)$ is sensitive to the internal structure of the operator and measures how densely the accessible region is populated by non-identity components. As a result, it is strongly affected by the presence or absence of coherent backscattering processes, leading to the observed non-monotonic dependence on disorder.}

%

\textcolor{red}{
	While both $\overline{m}(t)$ and $\overline{h}(t)$ exhibit similar logarithmic growth in the
MBL regime, they probe distinct aspects of operator dynamics. The operator
length $\overline{h}(t)$ characterizes the spatial extent of the operator and can be
interpreted as tracking the position of its propagation front. In contrast,
the operator mass $\overline{m}(t)$ measures the internal structure of the operator,
i.e., the number of non-identity components, and is therefore sensitive to
the degree of operator scrambling.
These two quantities can, in general, behave very differently. For example,
in non-interacting systems the operator front may propagate ballistically,
leading to a rapid growth of $\overline{h}(t)$, while the operator remains relatively
simple so that $\overline{m}(t)$ grows slowly.
In our setup, these observables are also constrained by construction:
the operator length $\overline{h}(t)$ sets an upper bound on the spatial support,
while the operator mass $\overline{m}(t)$ quantifies how densely this region is
populated by non-identity operators. In particular, $\overline{m}(t) \leq \overline{h}(t)$,
so that $h(t)$ characterizes the maximal extent of the operator, whereas
$\overline{m}(t)$ provides a measure of how much of this accessible region is
effectively occupied.
The fact that, in the present
case, both quantities display logarithmic growth reflects a nontrivial interplay between spatial spreading and
operator complexity in the MBL regime.
2}

\paragraph{Comparison with the $\ell$-bit model ---}
We conclude by commenting on the behavior of the operator length in the analytical $\ell$-bit model, as shown in Fig.~\ref{fig:lbit} (see~\ref{appendix:l-bit} for details). Once finite size effects are removed, the $\ell$-bit simulations display the same logarithmic behavior observed in the physical XXZ chain. The agreement in the behavior of $\overline{h}(t)$ between the XXZ model and the $\ell$-bit model suggests that -- for the system sizes and simulation times we have access to -- the slow operator dynamics is governed by exponentially decaying interactions between emergent localized degrees of freedom. Taken together, Figs.~\ref{fig:h_ED}, \ref{fig:mps_ent_h_m}, \ref{fig:sz_vs_delta}, \ref{logi:fig}, and \ref{fig:lbit} consistently demonstrate that:
\begin{itemize} 
  \item in the absence of interactions ($\Delta=0$) the operator remains strictly localized, and all observables saturate rapidly;
  \item in the interacting case ($\Delta=1$) the operator spreads logarithmically, producing slow growth in $\overline{h}(t)$, $\overline{m}(t)$, and $S_{\mathrm{op}}(t)$;
  \item time-averaged quantities collapse when rescaled by $\ln L$, in agreement with the $\ell$-bit prediction $\overline{h}(t)\sim \xi\ln t$; 
  \item the analytical results for the $\ell$-bit model reproduce the numerically observed logarithmic increase in time of $\overline{h}(t)$ of the interacting XXZ chain. 
\end{itemize} 
This combined numerical and analytical evidence demonstrates that the operator length, mass, and entanglement entropy provide coherent and sensitive probes of MBL operator dynamics, and neatly distinguish between Anderson localized and MBL regimes.
Furthermore, tensor-network simulations play a crucial role in this analysis. The MPS representation of the operator allows us to reach system  sizes and time scales far beyond the limits of ED, while  retaining full control over the operator structure. In the MBL regime, the slow growth of operator entanglement makes the MPO {\em Ansatz} particularly efficient: the linear increase in time of the bond dimension (being the operator entanglement entropy logarithmic in the bond dimension) mirrors the logarithmic spreading of the operator itself, enabling faithful simulations deep into the localized regime. 

\textcolor{red}{To conclude, the quantities $\overline{h}(t)$ and $\overline{m}(t)$ can be naturally related to the
light-cone structure probed by OTOCs.
The operator length $\overline{h}(t)$ tracks the position of the operator front,
and can be viewed as a coarse-grained measure of the maximal distance
at which the commutator $[\hat{O}(t), \hat{V}_j]$ becomes nonzero.
In this sense, $\overline{h}(t)$ provides a proxy for the OTOC light cone.
On the other hand, the operator mass $\overline{m}(t)$ quantifies the amount of
nontrivial operator weight within this region, thus probing the
internal buildup of operator complexity inside the light cone.
The fact that both quantities exhibit logarithmic growth in the MBL
regime is consistent with the logarithmic spreading of the OTOC front,
while providing complementary information about the structure of the
evolving operator. In the next section we are going to discuss strategies for possible experimental implementations.}


\section{Experimental considerations and strategies}
\label{sec:experiments}
The protocol we propose can be naturally adapted to several state-of-the-art quantum simulation platforms, including superconducting qubit processors, trapped-ion architectures, neutral atoms in optical tweezers, and programmable Rydberg-atom arrays~\cite{qi2019measuringoperatorsizegrowth, arute2019quantum, jurcevic2021demonstration, monroe2021programmable, guo2024site, barredo2016atom, manetsch2024tweezer, browaeys2020many, prxquantum2025_rydberg,levine2019_parallel,chiu2025_coherent,qiao2025_antiferro,miller2024_XYZ}. Here we do not focus on a specific hardware realization, but rather outline the minimal set of experimentally reasonable operations that are required to implement our scheme. The core ingredients are: (i) the preparation of a system--ancilla doubled state through the Choi--Jamiołkowski isomorphism~\cite{choi1975completely,jamiolkowski1972linear}; (ii) the controlled forward and backward time evolution under the many-body unitary \( \hat{U} \) and its inverse \( \hat{U}^{\dagger} \); and (iii) an efficient measurement strategy capable of reconstructing local expectation values of operators \( \hat{\mathcal{O}} \) from classical shadows~\cite{huang2020predicting,aaronson2018shadow}. Each of these components is routinely employed in modern quantum platforms and can be incorporated into a modular experimental protocol.

\paragraph{Time evolution and system--ancilla architecture ---}
The protocol relies on a doubled Hilbert space in which the system is accompanied by an ancillary copy. The ancillary qubits do not undergo the forward evolution; instead, they are evolved under \( \hat{U}^{\dagger} \). This asymmetry is essential, as it reproduces the Loschmidt-echo-type structure required for accessing out-of-time-order-like quantities and operator-spreading diagnostics. In practice, the implementation may use a second physical register, time-reversal techniques based on echo sequences, or synthetic dimensions that encode the ancilla. Since both \( \hat{U} \) and \( \hat{U}^{\dagger} \) can be realized via Trotter sequences of native gates (reversed in order), the experimental overhead remains comparable to that of standard dynamical experiments.

\paragraph{Local shadow reconstruction and the need for Bell measurements ---}

Instead of performing full tomography, which is unfeasible for large systems, we rely on classical-shadow techniques~\cite{huang2020predicting}. In this framework, a sequence of randomized measurements provides compressed yet unbiased estimators of the desired observables. In our case, however, the relevant operator acts on the composite Hilbert space of each system--ancilla pair. Therefore, the natural measurement basis is not the computational one, but the \textit{Bell basis}
\begin{equation}
\bigl\{ |\Phi^+\rangle, |\Phi^-\rangle, |\Psi^+\rangle, |\Psi^-\rangle \bigr\}
\end{equation}
on each site. Accessing this basis is crucial, as it allows one to extract the reduced shadow associated with a two-qubit subsystem, which in turn enables the reconstruction of expectation values of operators acting on the doubled space.

\paragraph{Realizing Bell measurements through local Clifford unitaries ---}

Direct Bell-state measurements are typically not native to most quantum hardware. Nevertheless, they can be implemented using standard tools by reducing them to single-qubit measurements preceded by a fixed Clifford transformation. This is the standard Bell-measurement reduction circuit introduced in Ref.~\cite{nielsen_chuang}.

For each system--ancilla pair \((\text{s}_i, \text{a}_i)\) we define the two-qubit Clifford unitary
\begin{equation}
    C_i = \mathrm{CNOT}_{\text{s}_i \rightarrow \text{a}_i}\, H_{\text{s}_i},
\end{equation}
in terms of the Hadamard gate $H_{\text{s}_i}$ acting on the system local qubit, and the CNOT gate acting in the local system-ancilla pair. This is the 
inverse of the conventional Bell-state preparation circuit. By construction,
\begin{equation}
C_i^\dagger \, |\Phi^+\rangle = |00\rangle, \quad
C_i^\dagger \, |\Phi^-\rangle = |01\rangle,\quad
C_i^\dagger \, |\Psi^+\rangle = |10\rangle,\quad
C_i^\dagger \, |\Psi^-\rangle =  |11\rangle.
\end{equation}
After this transformation, an effective Bell measurement is achieved by simply measuring each qubit in the \(\hat{\sigma}^z\) basis. This strategy is experimentally appealing because it uses only standard native gates (local Clifford unitaries and a two-qubit entangling gate) and single-qubit projective readout, which is routine in all leading quantum platforms.

\paragraph{Classical shadows from Bell-pair measurements ---}

Once the protocol is prepared, we distinguish two layers of Clifford operations. First, on each system qubit we apply a \emph{random local Clifford} $R_i$, and simultaneously on the corresponding ancilla qubit we apply the complex conjugate $R_i^*$. This joint operation, $(R_i \otimes R_i^*)$, preserves the correlations of the Choi state while introducing the randomness necessary for classical-shadow estimation~\cite{huang2020predicting}. In addition, while it may change the full Pauli spectrum of the operator, it keeps invariant the mass and length marginal probabilities we are interested in; in fact, it is easy to see that those random transformations cannot transform a local identity into a Pauli matrix and vice-versa.

Second, for each system--ancilla pair $(\text{s}_i, \text{a}_i)$ we apply the fixed Clifford $C_i$ that maps the Bell basis to the computational basis. After this transformation, a single experimental shot produces a bit-string
\begin{equation}
\{z_{\text{s}_i}, z_{\text{a}_i}\} \in \{0,1\}^2
\end{equation}
for each system--ancilla pair. Repeating the protocol for many shots yields a collection of classical snapshots. Each snapshot defines a classical estimator of a Pauli string in the Choi basis, namely $|\hat Q_{\text{shot}}\rangle\langle \hat Q_{\text{shot}}|$ where
\begin{equation}
\hat Q_{\text{shot}} = \bigotimes_{i=1}^L \hat P_i^{z_{\text{s}_i}, z_{\text{a}_i}}\,, \qquad \hat P_i^{z_{\text{s}_i}, z_{\text{a}_i}} \in \{ \mathbb{\hat I},\hat \sigma^x,\hat \sigma^y,\hat \sigma^z \}\,,
\end{equation}
which can be used to reconstruct expectation values of arbitrary local operators. Importantly, the marginal probabilities of interest, namely the extension
$h(\hat Q)$ of a Pauli string and its mass $m(\hat Q)$, can be estimated \emph{directly from these snapshots} without discarding any measurements. Specifically, for each snapshot, one computes $h(\hat Q_{\text{shot}})$,
$m(\hat Q_{\text{shot}})$, and increments the corresponding histogram bins.
In fact, since the marginal probabilities can be written as diagonal observables acting on the Choi state ({\it cfr.} Sec.~\ref{sec:MPS}),
namely $\hat P_{l} = 2^{-L}\sum_{\hat Q : h(\hat Q)=l} |\hat Q\rangle \langle \hat Q| $, and $\hat P_{m} = 2^{-L}\sum_{\hat Q : m(\hat Q)=m} |\hat Q\rangle \langle \hat Q| $, averaging over all snapshots then provides unbiased estimators for the marginal distributions
\begin{equation}
P(l) = \frac{1}{N_\text{shots}}\sum_{k=1}^{N_\text{shots}} \delta_{l, h(\hat Q_{\text{shot}}^{(k)})}\,, \qquad
P(m) = \frac{1}{N_\text{shots}}\sum_{k=1}^{N_\text{shots}} \delta_{m, m(\hat Q_{\text{shot}}^{(k)})}\,.
\end{equation}
Since the reconstruction acts pairwise on $(\text{s}_i,\text{a}_i)$, the statistical efficiency scales favorably with system size, and no global measurements are required. This procedure thus provides a practical and fully shot-efficient route to extract the operator marginals of interest.

\section{Conclusion}\label{sec:conclusion}
We have introduced the concept of \emph{operator length} as a complementary measure to the \emph{operator mass}. While the mass of a Pauli string counts the number of non-identity sites, and the mass of a general operator is obtained as its average over the Pauli decomposition, the operator length characterizes the spatial extent of the operator along a one-dimensional lattice. For a single Pauli string, it reduces to the position of the rightmost non-identity site, while for a generic operator it is averaged over the Pauli expansion weighted by the squared amplitudes. Both operator mass and length provide a way to quantify operator delocalization and the related scrambling of quantum information, a phenomenon strictly related to quantum-thermalization properties and crucial for complex quantum dynamics. While the operator mass is sensitive to the internal structure of the operator and measures how densely the accessible region inside a light cone of operator delocalization is populated by non-identity components, the operator length characterizes the spatial extent of the operator and can be interpreted as tracking the position of its propagation front.

A central contribution of this work is the first explicit procedure to compute the full probability distribution of operator length and mass within a matrix product operator (MPO) framework. This method allows exact, deterministic evaluation of these quantities, avoiding stochastic sampling, with a computational cost that scales polynomially rather than exponentially with the system size. This is in contrast to approaches based on full operator tomography to compute stabilizer Rényi entropies (SREs), which are generally exponentially expensive. Furthermore, the procedure we propose is experimentally feasible: by using the Choi-state representation of operators combined with the classical shadow approach, one can in principle reconstruct the relevant distributions in quantum simulators or other experimental platforms.

Applying these tools to the dynamics of an interacting disordered spin chain, we find that both the operator mass and operator length exhibit a logarithmic growth in time under strong disorder, reflecting the slow spreading of quantum information characteristic of many-body localized (MBL) systems. This behavior mirrors the logarithmic growth of entanglement entropy, indicating a close connection between operator spreading and entanglement growth. The logarithmic increase depends critically on interactions: in the non-interacting Anderson limit, both quantities quickly saturate to finite values, illustrating the essential role of many-body effects in sustaining slow delocalization. Our numerical findings are supported by a simple $\ell$-bit model, which reproduces the logarithmic growth via emergent local integrals of motion. For weaker disorder, the logarithmic growth is expected to crossover to faster, power-law-like spreading, consistent with thermalizing dynamics.

Looking forward, several directions are promising. One is the study of operator length and mass in monitored quantum systems, where measurements compete with unitary evolution and can induce entanglement transitions. Investigating the behavior of these quantities under monitored dynamics could shed light on the interplay between operator spreading, scrambling, and measurement-induced entanglement transitions. Moreover, since the operator-length and mass distributions are naturally defined in the Pauli basis, efficiently computable in MPO form, and experimentally accessible via Choi-state tomography and classical shadows, they provide a practical tool to probe operator delocalization in both simulations and experimental platforms.

Overall, the operator length, together with the operator mass, provides a versatile and physically transparent framework to quantify operator delocalization both in physical and in operator space. Our results demonstrate their utility in capturing slow logarithmic dynamics in interacting disordered systems, establish an efficient and exact MPO-based computational framework, and open new perspectives for studying operator spreading and information scrambling in a variety of quantum many-body settings.

\section{Acknowledgments}
The authors acknowledge valuable discussions with M.~Dalmonte, M.~Frau, and A.~Scocco. 
M.~C. acknowledges support from the PRIN2022 MUR project No.~2022R35ZBF - ``ManyQLowD''.
Part of the numerical simulations for this project were performed on the Ulysses v2 cluster at SISSA and on the Leonardo cluster at CINECA, through the SISSA–CINECA agreement for access to high-performance computing resources. A.~R. acknowledges computational resources from MUR, PON “Ricerca e Innovazione
2014-2020”, under Grant No. PIR01 00011 - (I.Bi.S.Co.). We acknowledge financial support from PNRR MUR Project PE0000023-NQSTI.
%
\appendix
\setcounter{figure}{0}
\renewcommand{\thefigure}{A\arabic{figure}}

\begin{figure}[t]
 \centering
 \includegraphics[width=100mm]{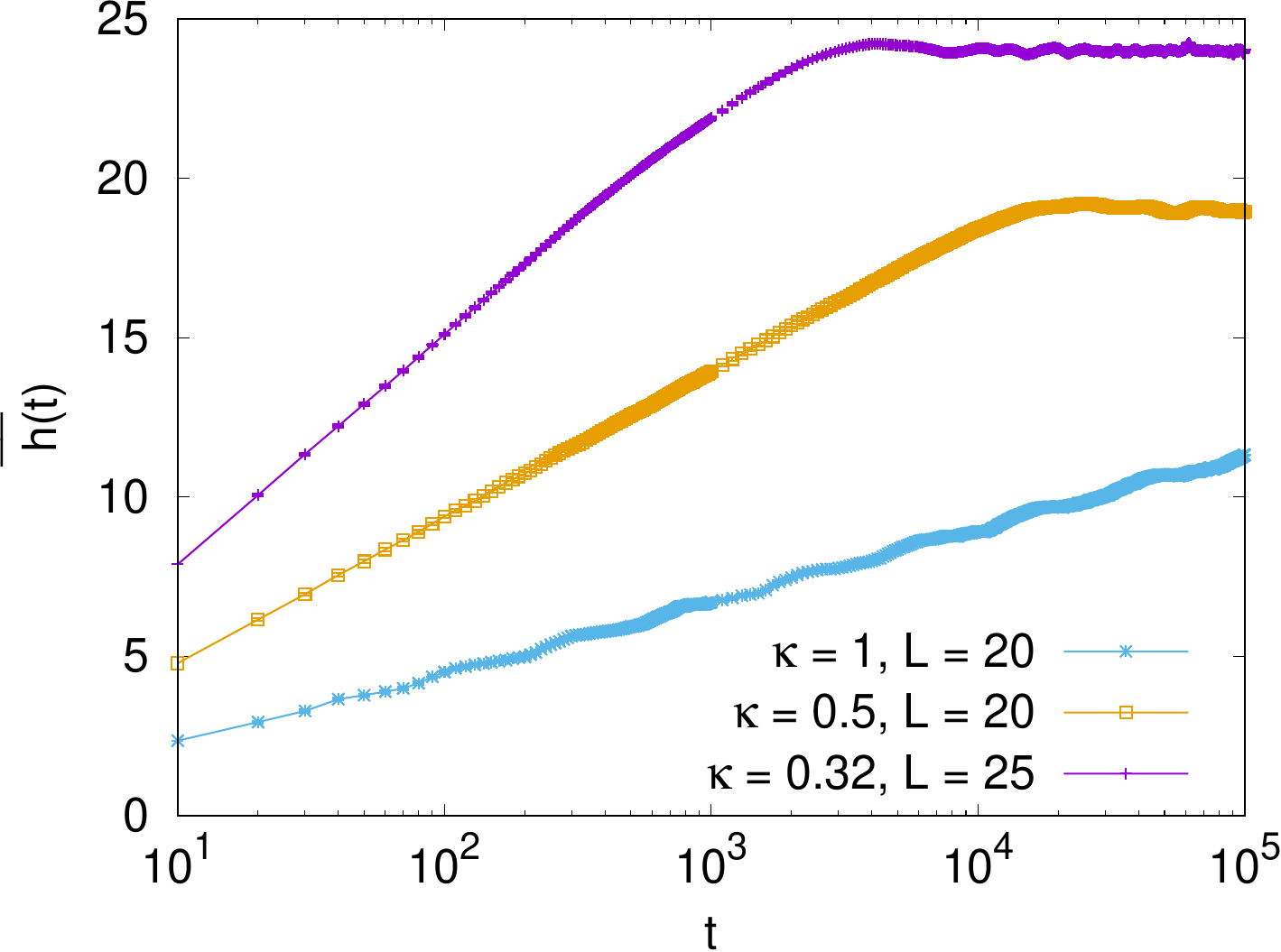}
 \caption{$\overline{h}(t)$ versus $t$ for the $\ell$-bit model of Eq.~\eqref{lbit:eqn} with $h_j=0$, for different values of the exponential decay rates $\kappa$ of the couplings $J_{jl}$, as defined in Eq.~\eqref{lbit:eqn} (see legend). Data are averaged over $N_{\rm r} = 192$ disorder realizations and we consider $W=1$.
 Notice the logarithmic scale on the horizontal axis. The logarithmic increase in time is visible until saturation to the finite-size boundary value ($h\leq L$) occurs. }
 \label{fig:lbit}
\end{figure}
\section{\texorpdfstring{$\ell$-bit model}{l-bit model}}\label{appendix:l-bit}
The logarithmic increase in time of $\overline{h}(t)$ in the strongly disordered interacting case can be analytically predicted using the $\ell$-bit model. This model with a superextensive number of localized integrals of motion has been introduced in Refs.~\cite{PhysRevLett.110.260601,PhysRevB.90.174202,PhysRevLett.111.127201,Nandkishore_2015} and demonstrated to describe the MBL phase of a quantum interacting system with strong disorder~\cite{Imbrie_2017}. It has been used to study different aspects of the MBL dynamics, pushing the calculations up to large times and system sizes, due to its simplicity. In particular, it has been applied to study the logarithmic entanglement entropy increase~\cite{PhysRevLett.110.260601}, the power-law decay of two-site entanglement~\cite{PhysRevB.94.214206}, the logarithmic light-cone of the OTOC~\cite{PhysRevB.95.060201,FAN2017707,chen2016universallogarithmicscramblingbody,Huang_2016}, the dynamical heterogeneity of the two-site entanglement~\cite{PhysRevB.105.184202}, and the ultraslow growth of number entropy~\cite{PhysRevLett.133.126502}. Its Hamiltonian is given by
\begin{equation}\label{lbit:eqn}
  \hat{H} = \sum_j h_j \hat{\tau}_j^z-\sum_{j\neq l} J_{jl} \, \hat{\tau}_j^z \, \hat{\tau}_l^z+\ldots\,, \qquad
  \mbox{with} \quad J_{jl} = W_{jl} \nep^{-\kappa|j-l|} \,,
\end{equation}
where $h_j$ are onsite random variables, and $W_{jl}$ are randomly and uniformly distributed in the interval $[-W,W]$. The $\hat{\tau}_j^\alpha$ are Pauli operator matrices (``$\ell$-bits'') given by superpositions of the original Pauli matrices $\hat{\sigma}_j^\alpha$ with amplitudes that decay exponentially, away from a given site. Therefore, each of the $\ell$-bit operators is localized around a given center -- with a localization length of order $\sim 1/\kappa$ -- , and for strong disorder has a huge overlap with a single one onsite $\hat{\sigma}_j^\alpha$ Pauli matrix. Notice the exponential decay  of the interactions  $J_{jl}$ with the distance between the localization centers, a phenomenological implementation of an analytically demonstrated result~\cite{imbrie1,PhysRevLett.111.127201,ROS2015420,Imbrie_2017}. We neglect further terms in the
Hamiltonian which comprise $n$-body interactions with $n \geq 3$, which is a controlled approximation for weakly interacting spins in the original microscopic model~\cite{PhysRevB.105.184202}. (The $n\geq 3$ terms provide higher-order corrections.) Here we fix also $h_j\equiv 0$ because onsite terms have been proved to have no action on the development of correlations and the logarithmic light-cone dynamics~\cite{PhysRevB.105.184202,PhysRevB.95.060201,FAN2017707,chen2016universallogarithmicscramblingbody,Huang_2016}. 

Let us study the delocalization in time of the initially localized operator
\begin{equation}
  \hat{\mathcal{O}} = \hat{\tau}_1^x\otimes \hat{\mathbb{I}}_2\otimes \cdots \otimes \hat{\mathbb{I}}_L\,.
\end{equation}
This operator is localized near the leftmost edge of the chain, with a support of the order of $1/\kappa$. The Heisenberg dynamics of this operator can be found in an exact analytical way, studying the evolution of the operators $\hat{\tau}_1^x$, $\hat{\tau}_1^y$. The Heisenberg equations $\tfrac{d}{dt} \hat{\tau}_1^{x/y}(t) = i [\hat{H},\hat{\tau}_1^{x/y}]$ for the two operators can be obtained susbstituting Eq.~\eqref{lbit:eqn} and evaluating the commutators using the commutation relations $[\hat{\tau}_j^\alpha,\hat{\tau}_l^\beta] = \epsilon^{\alpha\beta\gamma}\delta_{jl}\hat{\tau}_j^\gamma$
\begin{equation}
  \frac{d}{dt} \hat{\tau}_1^x(t) = -2\sum_{j\neq 1} J_{j1}\hat{\tau}_j^z\hat{\tau}_1^y\,, \qquad
  \frac{d}{dt} \hat{\tau}_1^y(t) = 2\sum_{j\neq 1} J_{j1}\hat{\tau}_j^z\hat{\tau}_1^x\,.
\end{equation}
These equations can be analytically solved and one finds
\begin{align}\label{opertime:eqn}
  \hat{\tau}_1^x(t) &= \hat{\tau}_1^x \cos\bigg( 2\sum_{j\neq 1} J_{j1}\hat{\tau}_j^z t \bigg) - \hat{\tau}_1^y \sin \bigg(2\sum_{j\neq 1} J_{j1}\hat{\tau}_j^z t\bigg)\,,\nonumber\\
  \hat{\tau}_1^y(t) &= \hat{\tau}_1^y \cos\bigg(2\sum_{j\neq 1} J_{j1}\hat{\tau}_j^z t\bigg) + \hat{\tau}_1^x \sin\bigg(2\sum_{j\neq 1} J_{j1}\hat{\tau}_j^z t\bigg)\,.
\end{align}
We now expand the sine and cosine terms, such that
\begin{align}
  &\cos\bigg(2\sum_{j\neq 1} J_{jl}\hat{\tau}_j^z t\bigg)  = \frac12 \bigg[ \exp \bigg(2i\sum_{j\neq 1} J_{j1}\hat{\tau}_j^z t\bigg)+\exp\bigg(-2i\sum_{j\neq 1} J_{j1}\hat{\tau}_j^z t\bigg) \bigg]\nonumber\\
   &=\frac12\bigg[\prod_{j\neq 1} \exp\left(2iJ_{j1}\hat{\tau}_j^z t\right)+\prod_{j\neq 1}\exp\left(-2i J_{j1}\hat{\tau}_j^z t\right)\bigg]\nonumber\\
   &=\frac12\bigg[\prod_{j\neq 1} \left[\cos\left(2 J_{j1}t\right)+i\hat{\tau}_j^z\sin\left(2 J_{j1}t\right)\right]+\prod_{j\neq 1}\left[\cos\left(2 J_{j1}t\right)-i\hat{\tau}_j^z\sin\left(2 J_{j1}t\right)\right]\bigg]
\end{align}
and
\begin{align}
  &\sin\bigg(2\sum_{j\neq 1} J_{jl}\hat{\tau}_j^z t\bigg)  = \frac{1}{2i}\bigg[\exp\bigg(2i\sum_{j\neq 1} J_{j1}\hat{\tau}_j^z t\bigg)-\exp\bigg(-2i\sum_{j\neq 1} J_{j1}\hat{\tau}_j^z t\bigg)\bigg]\nonumber\\
   &=\frac{1}{2i}\bigg[\prod_{j\neq 1} \Big[\cos\left(2 J_{j1}t\right)+i\hat{\tau}_j^z\sin\left(2 J_{j1}t\right)\Big]\!-\!\prod_{j\neq 1}\Big[ \cos\left(2 J_{j1}t\right)-i\hat{\tau}_j^z\sin\left(2 J_{j1}t\right)\Big] \bigg].
\end{align}
This expansion is useful to evaluate the Hilbert-Schmidt inner products $\mathcal{A}_Q=\Tr[\hat{Q}\hat{\tau}_1^x]$ in Eq.~\eqref{opertime:eqn}, considering the $\ell$-bit Pauli string operators
\begin{equation}
  \hat{Q} = \hat{\tau}_1^{\alpha_1}\otimes\hat{\tau}_2^{\alpha_2}\otimes\cdots\otimes\hat{\tau}_L^{\alpha_L}\,.
\end{equation}
We can use $\ell-bit$ Pauli strings in stead of the original ones due to the above-discussed physical-space localization in the $\hat{\sigma}_j^{\alpha_j}$ of the $\tau_l^{\alpha_l}$ operators. First of all, we notice that one can have a nontrivial Hilbert-Schmidt inner product only with Pauli operators that have $\hat{\tau}_1^x$ or $\hat{\tau}_1^y$ on site 1 and the identity or $\hat{\tau}_j^z$ for $j\neq 1$. There are $2^L$ of such operators, much less than the $4^L$ of the entire Pauli set. There are two subsets of Pauli operators that are worth being considered separately:
\begin{itemize}
\item The first one is given by operators as
\begin{equation}
  \hat{Q} = \hat{\tau}_1^x\prod_{j=2}^L\hat{\tau}_j^{\alpha_j},\quad{\rm with}\quad \alpha_j = 0,\, 3 \,,
\end{equation}
(with $\alpha_j=0$ for the identity and $\alpha_j=3$ for $\hat{\tau}_j^z$). For this we get
\begin{align}
  \mathcal{A}_Q(t)=\Tr \big[ \hat{Q} \hat{\tau}_1^x(t) \big] = \; & 2^{L-1} \Bigg\{ \prod_{j=2}^L \Big[ \delta_{\alpha_j0}\cos\left(2 J_{j1}t\right)+i\delta_{\alpha_j3}\sin\left(2 J_{j1}t\right) \Big] \nonumber\\
  & \qquad + \prod_{j=2}^L \Big[ \delta_{\alpha_j0}\cos\left(2 J_{j1}t\right)-i\delta_{\alpha_j3}\sin\left(2 J_{j1}t\right) \Big] \Bigg\}\,,
\end{align}
so we can recast it as
\begin{align}
  \mathcal{A}_Q(t) = \; & 2^{L-1} \!\! \prod_{j>1\,\text{s.\,t.}\, \alpha_j = 0} \cos\left(2 J_{j1}t\right) \nonumber \\
  &\times \Bigg\{ \prod_{j'>1\,\text{s.\,t.} \alpha_{j'} = 3} \!\! i\sin\left(2 J_{j'1}t\right)+ \!\! \prod_{j'>1\,\text{s.\,t.}\, \alpha_{j'} = 3}(-i)\sin\left(2 J_{j'1}t\right)\Bigg\}\,.
\end{align}
Therefore, this quantity is nonvanishing only if the number of sites where $\alpha_{j'} = 3$ is {\em even}. If this condition is verified we get
\begin{equation}\label{qal:eqn}
  |\mathcal{A}_Q(t)|^2=4^{L}\prod_{j>1\,\text{s.\,t.}\, \alpha_j = 0} \cos^2\left(2 J_{j1}t\right)\prod_{j'>1\,\text{s.\,t.} \alpha_{j'} = 3}\sin^2\left(2 J_{j'1}t\right)\,.
\end{equation}
\item The other subset of Pauli operators is given by
\begin{equation}
  \hat{P} = \hat{\tau}_1^y\prod_{j=2}^L\hat{\tau}_j^{\alpha_j}\quad{\rm with}\quad \alpha_j = 0,\, 3\,.
\end{equation}
They provide inner-product contributions of the form
\begin{align}
  \mathcal{A}_P(t)=\Tr \big[ \hat{P} \hat{\tau}_1^x(t) \big] = \; & -2^{L-1}i\Bigg\{\prod_{j=2}^L \Big[ \delta_{\alpha_j0}\cos\left(2 J_{j1}t\right)+i\delta_{\alpha_j3}\sin\left(2 J_{j1}t\right) \Big] \nonumber\\
  &\hspace{1.25cm} -\prod_{j=2}^L \Big[ \delta_{\alpha_j0}\cos\left(2 J_{j1}t\right)-i\delta_{\alpha_j3}\sin\left(2 J_{j1}t\right) \Big] \Bigg\}\,,
\end{align}
that can be recast as
\begin{align}
  \mathcal{A}_P(t) = \; & -2^{L-1}i\prod_{j>1\,\text{s.\,t.}\, \alpha_j = 0} \cos\left(2 J_{j1}t\right) \nonumber \\
  & \times \! \Bigg\{\prod_{j'>1\,\text{s.\,t.} \alpha_{j'} = 3}i\sin\left(2 J_{j'1}t\right)-\!\!\prod_{j'>1\,\text{s.\,t.}\, \alpha_{j'} = 3} \!\! (-i)\sin\left(2 J_{j'1}t\right)\Bigg\}.
\end{align}
In this case, we get a nonvanishing value only when the number of sites where $\alpha_{j'}=3$ is {\em odd}. When this condition is verified, we get
\begin{equation}\label{pbl:eqn}
  |\mathcal{A}_P(t)|^2= 4^L\prod_{j>1\,\text{s.\,t.}\, \alpha_j = 0} \cos^2\left(2 J_{j1}t\right)\prod_{j'>1\,\text{s.\,t.} \alpha_{j'} = 3}\sin^2\left(2 J_{j'1}t\right)\,.
\end{equation}
\end{itemize}
Comparing Eqs.~\eqref{qal:eqn} and~\eqref{pbl:eqn}, we see that both subsets of operators provide modulus-square inner products given by the same analytical formula. We can insert Eqs.~\eqref{qal:eqn} and~\eqref{pbl:eqn} into the expression for $h(t)$ given by Eq.~\eqref{hot:eqn}, and numerically evaluate the delocalization length $\overline{h}(t)$ averaged over realizations. Some examples of $\overline{h}(t)$ versus $t$ are shown in Fig.~\ref{fig:lbit}: one can see a quite clear logarithmic increase in time until the finite-size boundary value of $h$ is reached. (For finite size there is the constraint $h\leq L$) We find that the slope of the logarithmic increase decreases with $1/\kappa$. By applying a least-square linear fit to $\overline{h}(t)$ versus $\ln t$ we find the following values of the slope
\begin{center}
\begin{tabular}{|c|c|}
\hline
$\kappa$&Slope\\
\hline
1.0 & $0.98\pm 0.01$\\
0.5& $1.98 \pm 0.001$\\
0.32 & $ 3.1537\pm 0.004$\\
\hline
\end{tabular}
\end{center}
One sees that the slope decreases with increasing $\kappa$, approximately as $\sim 1/\kappa$. The reason is that for smaller $\kappa$ there is a stronger coupling between the $\ell$-bits (the decay length of an $\ell$-bit in physical space is larger) and so correlations can propagate more efficiently, but anyway at a logarithmic rate. The logarithmic increase of $\overline{h}(t)$ we find here is in full agreement with the numerical results shown in Sec.~\ref{sec:results}.

\bibliography{biblio}
\end{document}